\documentclass[floats,floatfix,amssymb,prl,twocolumn,superscriptaddress,nofootinbib]{revtex4-2}

\usepackage{subcaption}
\usepackage{ragged2e}
\DeclareCaptionJustification{justified}{\justifying}
\makeatletter
\newcommand{\subsetsim}{\mathrel{\mathpalette\subset@sim\relax}}
\newcommand{\subset@sim}[2]{%
  \vtop{\offinterlineskip\m@th
    \ialign{\hfil##\cr
      $#1\subset$\cr\noalign{\kern0.5pt}\scalebox{0.9}{$#1\sim$}\cr
    }%
  }%
}
\makeatother

\usepackage{amssymb,amsmath,verbatim,mathtools,needspace,enumitem,etoolbox,graphicx,physics,microtype,afterpage,bm}
\usepackage[dvipsnames, usenames]{xcolor}
\definecolor{linkcolor}{rgb}{0.0,0.3,0.5}
\usepackage{booktabs}

\definecolor{oucrimsonred}{rgb}{0.6, 0.0, 0.0}
\definecolor{persianblue}{rgb}{0.11, 0.22, 0.73}
\definecolor{forestgreen}{rgb}{0.13,0.35,0.13}
\usepackage[unicode, 
colorlinks=true, 
linkcolor=persianblue, 
citecolor=forestgreen, 
filecolor=persianblue,
urlcolor=oucrimsonred, 
pdfusetitle]{hyperref}

\usepackage[all]{hypcap}
\usepackage[T1]{fontenc}
\usepackage[utf8]{inputenc}
\usepackage{tabularx}
\usepackage{adjustbox}
\usepackage{float}
\usepackage{ulem}
\usepackage{xfrac}
\usepackage{orcidlink}

\definecolor{harvardcrimson}{rgb}{0.79, 0.0, 0.09}

\definecolor{azure}{rgb}{0.0, 0.5, 1.0}
\definecolor{deepfuchsia}{rgb}{0.76, 0.33, 0.76}
\definecolor{VioletRed4}{rgb}{0.55, 0.13, .32}
 
\definecolor{harvardcrimson}{rgb}{0.79, 0.0, 0.09}
\definecolor{oceanboatblue}{rgb}{0.0, 0.47, 0.75}
\definecolor{persianblue}{rgb}{0.11, 0.22, 0.73}
\definecolor{egyptianblue}{rgb}{0.06, 0.2, 0.65}
\definecolor{navyblue}{rgb}{0.0, 0.0, 0.5}

\definecolor{verdechiaro}{rgb}{0.6,1,0.6}
\definecolor{giallochiaro}{rgb}{1,1,0.6}
\definecolor{bluscuro}{rgb}{0.15, 0.2, 0.9}
\definecolor{verdes}{rgb}{0.1, 0.5, 0.1}%
\definecolor{tangerineyellow}{rgb}{1.0, 0.8, 0.0}

\usepackage{multirow}
\usepackage{pifont}
\usepackage{fontawesome}
\usepackage{lmodern}

\usepackage{multirow}

\allowdisplaybreaks
\usepackage{tikz}
\usepackage{tikz-feynman}
\usetikzlibrary{tikzmark}
\usepackage{tcolorbox}
\usepackage{pifont}
\usepackage{color}
\usepackage{framed}

\definecolor{rossos}{cmyk}{0,1,1,0.55}
\definecolor{bluscuro}{rgb}{0.15, 0.2, .85}
\definecolor{bluchiaro}{cmyk}{1,.3,0.,0.1}
\definecolor{ForestGreen}{rgb}{0.13, 0.55, 0.13}

\newcommand{\MPl}{\bar{M}_{\textrm{\tiny{Pl}}}}

\newtcolorbox{mybox}{colback=mycolor!5!white,colframe=azure!75!black}

\def\nn{\nonumber}

\def\bea{\begin{eqnarray}}
\def\eea{\end{eqnarray}}

\newcommand{\bs}{\begin{subequations}}
\newcommand{\es}{\end{subequations}}

\newcommand{\be}{\begin{equation}}
\newcommand{\ee}{\end{equation}}

\def\lsim{\mathrel{\rlap{\lower4pt\hbox{\hskip0.5pt$\sim$}}
    \raise1pt\hbox{$<$}}}         
\def\gsim{\mathrel{\rlap{\lower4pt\hbox{\hskip0.5pt$\sim$}}
    \raise1pt\hbox{$>$}}}         

\makeatletter
\def\l@subsubsection#1#2{}
\makeatother

\newcommand{\sapienza}{Dipartimento di Fisica, Sapienza Università 
	di Roma, Piazzale Aldo Moro 5, 00185, Roma, Italy}
\newcommand{\infn}{INFN Roma\,1, Piazzale Aldo Moro 2, 00185, Roma, Italy}
\newcommand{\jhu}{William H.\ Miller III Department of Physics and Astronomy, Johns Hopkins University, \\ 3400 N. Charles Street, Baltimore, Maryland, 21218, USA}

\begin{document}

\title{
Insights into the highest natural scale: \\
Finite naturalness challenges inflationary dynamics
}

\author{Pier Giuseppe Catinari\orcidlink{0009-0008-5416-1770}}
\email{piergiuseppe.catinari@uniroma1.it}
\affiliation{\sapienza}
\affiliation{\infn}

\author{Loris Del Grosso\hspace{0.05cm}\orcidlink{0000-0002-6722-4629}}
\email{ldelgro1@jh.edu}
\affiliation{\jhu}

\author{Luca Di Giovanni}
\email{digiovanni.1395475@studenti.uniroma1.it}
\affiliation{\sapienza}

\author{Alfredo Urbano\orcidlink{0000-0002-0488-3256}}
\email{alfredo.urbano@uniroma1.it}
\affiliation{\sapienza}
\affiliation{\infn}

\date{\today}

\begin{abstract}
We apply the criterion of finite naturalness to the limiting case of a generic heavy sector decoupled from the Standard Model. The sole and unavoidable exception to this decoupling arises from gravitational interactions. We demonstrate that gravity can couple the Higgs to the heavy scale significantly earlier than the well-known three-loop top-quark-mediated diagrams discussed in previous literature. As an application, we show that finite naturalness disfavors large-field inflationary models involving super-Planckian field excursions. In contrast, in the small-field regime, achieving successful inflation requires substantial fine-tuning of the initial conditions, in agreement with previous results. Recent data from the Atacama Cosmology Telescope further amplify the tension between naturalness and fine-tuning, challenging the theoretical robustness of single-field inflation as a compelling explanation for the origin of the universe.
\end{abstract}

\maketitle

{
  \hypersetup{linkcolor=black}
}

\normalem

 \textit{\textbf{Introduction.}} Whether the Standard Model (SM) is a \textit{natural} theory or not remains an open question—one that has fueled remarkably theoretical development (see e.g. the review~\cite{Craig:2022eqo} and the references cited therein). In a nutshell, a theory is \textit{natural} if the contribution to the Higgs mass $\delta m_h$ arising from quantum corrections is not larger than Higgs pole mass $M_h \approx 125 \, {\rm GeV}$. 
However, it is less clear how to implement the naturalness principle concretely. For example, if we assume some unknown scale of new physics $\Lambda_{\rm NP}$ and compute the one-loop correction to the Higgs mass coming from the top quark using cutoff regularization, we would get  $\delta m_h^2 \sim y_t^2/(4\pi)^2\Lambda_{\rm NP}^2 $, where $y_t \approx 1$ is the top Yukawa coupling. The condition $\delta m_h^2  \lesssim M_h^2$ would imply $\Lambda_{\rm NP} \lesssim \rm{TeV}$. However,  no new physics has been so far seen at LHC with $\sqrt{s} \approx 14 \, {\rm TeV}$.
This argument invests the regulator with a physical meaning~\cite{RevModPhys.55.583, Branchina:2022gll}, namely $\Lambda_{\rm{NP}}$ is the scale where the Higgs mass becomes calculable. Despite this physical interpretation, quadratic divergences are regulator-dependent artifacts and do not manifest in all regularization schemes—for example, they are absent in dimensional regularization, where power-law divergences are systematically set to zero.
\begin{figure}[!t]

\tikzset{every picture/.style={line width=0.85pt}} 

\begin{tikzpicture}[scale = 0.95, x=0.6pt,y=0.6pt,yscale=-1,xscale=1]
	
	\draw    (149,480.4) -- (149,60.6) ;
	\draw [shift={(149,57.6)}, rotate = 90] [fill={rgb, 255:red, 0; green, 0; blue, 0 }  ][line width=0.08]  [draw opacity=0] (8.93,-4.29) -- (0,0) -- (8.93,4.29) -- cycle    ;
	\draw  [dash pattern={on 4.5pt off 4.5pt}]  (200.7,233.5) -- (230.65,233.5) ;
	\draw  [dash pattern={on 4.5pt off 4.5pt}]  (269.7,233.5) -- (299.65,233.5) ;
	\draw  [draw opacity=0][line width=1.5]  (267.75,233.81) .. controls (267.58,243.91) and (259.34,252.05) .. (249.2,252.05) .. controls (239.01,252.05) and (230.74,243.83) .. (230.65,233.66) -- (249.2,233.5) -- cycle ; \draw  [color={rgb, 255:red, 0; green, 118; blue, 255 }  ,draw opacity=1 ][line width=1.5]  (267.75,233.81) .. controls (267.58,243.91) and (259.34,252.05) .. (249.2,252.05) .. controls (239.01,252.05) and (230.74,243.83) .. (230.65,233.66) ;  
	\draw  [fill={rgb, 255:red, 0; green, 0; blue, 0 }  ,fill opacity=1 ] (266.18,233.5) .. controls (266.18,232.63) and (266.88,231.93) .. (267.75,231.93) .. controls (268.62,231.93) and (269.32,232.63) .. (269.32,233.5) .. controls (269.32,234.37) and (268.62,235.07) .. (267.75,235.07) .. controls (266.88,235.07) and (266.18,234.37) .. (266.18,233.5) -- cycle ;
	\draw  [fill={rgb, 255:red, 0; green, 0; blue, 0 }  ,fill opacity=1 ] (229.08,233.5) .. controls (229.08,232.63) and (229.78,231.93) .. (230.65,231.93) .. controls (231.52,231.93) and (232.22,232.63) .. (232.22,233.5) .. controls (232.22,234.37) and (231.52,235.07) .. (230.65,235.07) .. controls (229.78,235.07) and (229.08,234.37) .. (229.08,233.5) -- cycle ;
	\draw  [dash pattern={on 4.5pt off 4.5pt}]  (200.6,349.8) -- (223.55,349.8) ;
	\draw  [dash pattern={on 4.5pt off 4.5pt}]  (276.27,349.8) -- (300.22,349.8) ;
	\draw   (223.55,349.8) .. controls (223.55,335.24) and (235.35,323.44) .. (249.91,323.44) .. controls (264.47,323.44) and (276.27,335.24) .. (276.27,349.8) .. controls (276.27,364.36) and (264.47,376.16) .. (249.91,376.16) .. controls (235.35,376.16) and (223.55,364.36) .. (223.55,349.8) -- cycle ;
	\draw   (248.91,341.22) .. controls (248.22,340.59) and (247.68,339.4) .. (247.72,337.49) .. controls (247.82,332.49) and (251.7,332.57) .. (251.66,334.57) .. controls (251.62,336.57) and (247.74,336.49) .. (247.84,331.49) .. controls (247.95,326.49) and (251.83,326.57) .. (251.79,328.57) .. controls (251.74,330.57) and (247.87,330.49) .. (247.97,325.49) .. controls (247.98,324.9) and (248.05,324.38) .. (248.15,323.92) ;
	\draw  [color={rgb, 255:red, 0; green, 118; blue, 255 }  ,draw opacity=1 ][line width=1.5]  (240.62,349.8) .. controls (240.62,344.67) and (244.78,340.51) .. (249.91,340.51) .. controls (255.04,340.51) and (259.2,344.67) .. (259.2,349.8) .. controls (259.2,354.93) and (255.04,359.09) .. (249.91,359.09) .. controls (244.78,359.09) and (240.62,354.93) .. (240.62,349.8) -- cycle ;
	\draw   (248.91,376.22) .. controls (248.22,375.59) and (247.68,374.4) .. (247.72,372.49) .. controls (247.82,367.49) and (251.7,367.57) .. (251.66,369.57) .. controls (251.62,371.57) and (247.74,371.49) .. (247.84,366.49) .. controls (247.95,361.49) and (251.83,361.57) .. (251.79,363.57) .. controls (251.74,365.57) and (247.87,365.49) .. (247.97,360.49) .. controls (247.98,359.9) and (248.05,359.38) .. (248.15,358.92) ;
	\draw  [fill={rgb, 255:red, 0; green, 0; blue, 0 }  ,fill opacity=1 ] (248.33,323.44) .. controls (248.33,322.57) and (249.04,321.87) .. (249.91,321.87) .. controls (250.78,321.87) and (251.48,322.57) .. (251.48,323.44) .. controls (251.48,324.31) and (250.78,325.02) .. (249.91,325.02) .. controls (249.04,325.02) and (248.33,324.31) .. (248.33,323.44) -- cycle ;
	\draw  [fill={rgb, 255:red, 0; green, 0; blue, 0 }  ,fill opacity=1 ] (248.33,340.51) .. controls (248.33,339.64) and (249.04,338.94) .. (249.91,338.94) .. controls (250.78,338.94) and (251.48,339.64) .. (251.48,340.51) .. controls (251.48,341.38) and (250.78,342.09) .. (249.91,342.09) .. controls (249.04,342.09) and (248.33,341.38) .. (248.33,340.51) -- cycle ;
	\draw  [fill={rgb, 255:red, 0; green, 0; blue, 0 }  ,fill opacity=1 ] (274.69,349.8) .. controls (274.69,348.93) and (275.4,348.23) .. (276.27,348.23) .. controls (277.14,348.23) and (277.84,348.93) .. (277.84,349.8) .. controls (277.84,350.67) and (277.14,351.37) .. (276.27,351.37) .. controls (275.4,351.37) and (274.69,350.67) .. (274.69,349.8) -- cycle ;
	\draw  [fill={rgb, 255:red, 0; green, 0; blue, 0 }  ,fill opacity=1 ] (221.98,349.8) .. controls (221.98,348.93) and (222.68,348.23) .. (223.55,348.23) .. controls (224.42,348.23) and (225.12,348.93) .. (225.12,349.8) .. controls (225.12,350.67) and (224.42,351.37) .. (223.55,351.37) .. controls (222.68,351.37) and (221.98,350.67) .. (221.98,349.8) -- cycle ;
	\draw  [fill={rgb, 255:red, 0; green, 0; blue, 0 }  ,fill opacity=1 ] (248.33,359.09) .. controls (248.33,358.22) and (249.04,357.51) .. (249.91,357.51) .. controls (250.78,357.51) and (251.48,358.22) .. (251.48,359.09) .. controls (251.48,359.96) and (250.78,360.66) .. (249.91,360.66) .. controls (249.04,360.66) and (248.33,359.96) .. (248.33,359.09) -- cycle ;
	\draw  [fill={rgb, 255:red, 0; green, 0; blue, 0 }  ,fill opacity=1 ] (248.33,376.16) .. controls (248.33,375.29) and (249.04,374.58) .. (249.91,374.58) .. controls (250.78,374.58) and (251.48,375.29) .. (251.48,376.16) .. controls (251.48,377.03) and (250.78,377.73) .. (249.91,377.73) .. controls (249.04,377.73) and (248.33,377.03) .. (248.33,376.16) -- cycle ;
	\draw  [dash pattern={on 4.5pt off 4.5pt}]  (200.8,94.2) -- (223.75,94.2) ;
	\draw  [dash pattern={on 4.5pt off 4.5pt}]  (276.47,94.2) -- (300.42,94.2) ;
	\draw   (223.75,94.2) .. controls (223.75,79.64) and (235.55,67.84) .. (250.11,67.84) .. controls (264.67,67.84) and (276.47,79.64) .. (276.47,94.2) .. controls (276.47,108.75) and (264.67,120.55) .. (250.11,120.55) .. controls (235.55,120.55) and (223.75,108.75) .. (223.75,94.2) -- cycle ;
	\draw  [color={rgb, 255:red, 0; green, 120; blue, 255 }  ,draw opacity=1 ][line width=1.5]  (240.82,94.2) .. controls (240.82,89.07) and (244.98,84.91) .. (250.11,84.91) .. controls (255.24,84.91) and (259.4,89.07) .. (259.4,94.2) .. controls (259.4,99.32) and (255.24,103.48) .. (250.11,103.48) .. controls (244.98,103.48) and (240.82,99.32) .. (240.82,94.2) -- cycle ;
	\draw    (251.61,67.84) .. controls (253.28,69.51) and (253.28,71.17) .. (251.61,72.84) .. controls (249.94,74.51) and (249.94,76.17) .. (251.61,77.84) .. controls (253.28,79.51) and (253.28,81.17) .. (251.61,82.84) -- (251.61,84.91) -- (251.61,84.91)(248.61,67.84) .. controls (250.28,69.51) and (250.28,71.17) .. (248.61,72.84) .. controls (246.94,74.51) and (246.94,76.17) .. (248.61,77.84) .. controls (250.28,79.51) and (250.28,81.17) .. (248.61,82.84) -- (248.61,84.91) -- (248.61,84.91) ;
	\draw    (251.61,103.48) .. controls (253.28,105.15) and (253.28,106.81) .. (251.61,108.48) .. controls (249.94,110.15) and (249.94,111.81) .. (251.61,113.48) .. controls (253.28,115.15) and (253.28,116.81) .. (251.61,118.48) -- (251.61,120.55) -- (251.61,120.55)(248.61,103.48) .. controls (250.28,105.15) and (250.28,106.81) .. (248.61,108.48) .. controls (246.94,110.15) and (246.94,111.81) .. (248.61,113.48) .. controls (250.28,115.15) and (250.28,116.81) .. (248.61,118.48) -- (248.61,120.55) -- (248.61,120.55) ;
	\draw  [fill={rgb, 255:red, 0; green, 0; blue, 0 }  ,fill opacity=1 ] (248.53,67.84) .. controls (248.53,66.97) and (249.24,66.26) .. (250.11,66.26) .. controls (250.98,66.26) and (251.68,66.97) .. (251.68,67.84) .. controls (251.68,68.71) and (250.98,69.41) .. (250.11,69.41) .. controls (249.24,69.41) and (248.53,68.71) .. (248.53,67.84) -- cycle ;
	\draw  [fill={rgb, 255:red, 0; green, 0; blue, 0 }  ,fill opacity=1 ] (274.89,94.2) .. controls (274.89,93.33) and (275.6,92.62) .. (276.47,92.62) .. controls (277.34,92.62) and (278.04,93.33) .. (278.04,94.2) .. controls (278.04,95.06) and (277.34,95.77) .. (276.47,95.77) .. controls (275.6,95.77) and (274.89,95.06) .. (274.89,94.2) -- cycle ;
	\draw  [fill={rgb, 255:red, 0; green, 0; blue, 0 }  ,fill opacity=1 ] (248.53,120.55) .. controls (248.53,119.68) and (249.24,118.98) .. (250.11,118.98) .. controls (250.98,118.98) and (251.68,119.68) .. (251.68,120.55) .. controls (251.68,121.42) and (250.98,122.13) .. (250.11,122.13) .. controls (249.24,122.13) and (248.53,121.42) .. (248.53,120.55) -- cycle ;
	\draw  [fill={rgb, 255:red, 0; green, 0; blue, 0 }  ,fill opacity=1 ] (222.18,94.2) .. controls (222.18,93.33) and (222.88,92.62) .. (223.75,92.62) .. controls (224.62,92.62) and (225.32,93.33) .. (225.32,94.2) .. controls (225.32,95.06) and (224.62,95.77) .. (223.75,95.77) .. controls (222.88,95.77) and (222.18,95.06) .. (222.18,94.2) -- cycle ;
	\draw  [fill={rgb, 255:red, 0; green, 0; blue, 0 }  ,fill opacity=1 ] (248.53,103.48) .. controls (248.53,102.61) and (249.24,101.91) .. (250.11,101.91) .. controls (250.98,101.91) and (251.68,102.61) .. (251.68,103.48) .. controls (251.68,104.35) and (250.98,105.06) .. (250.11,105.06) .. controls (249.24,105.06) and (248.53,104.35) .. (248.53,103.48) -- cycle ;
	\draw  [fill={rgb, 255:red, 0; green, 0; blue, 0 }  ,fill opacity=1 ] (248.53,84.91) .. controls (248.53,84.04) and (249.24,83.33) .. (250.11,83.33) .. controls (250.98,83.33) and (251.68,84.04) .. (251.68,84.91) .. controls (251.68,85.78) and (250.98,86.48) .. (250.11,86.48) .. controls (249.24,86.48) and (248.53,85.78) .. (248.53,84.91) -- cycle ;
	\draw  [dash pattern={on 4.5pt off 4.5pt}]  (200.28,188.6) -- (300.14,188.6) ;
	\draw  [color={rgb, 255:red, 0; green, 117; blue, 255 }  ,draw opacity=1 ][line width=1.5]  (234.92,172.81) .. controls (234.92,164.09) and (241.99,157.02) .. (250.71,157.02) .. controls (259.43,157.02) and (266.5,164.09) .. (266.5,172.81) .. controls (266.5,181.53) and (259.43,188.6) .. (250.71,188.6) .. controls (241.99,188.6) and (234.92,181.53) .. (234.92,172.81) -- cycle ;
	\draw    (138.33,349.77) -- (158.33,349.8) ;
	\draw  [dash pattern={on 4.5pt off 4.5pt}]  (200.37,288.9) -- (230.32,288.9) ;
	\draw  [dash pattern={on 4.5pt off 4.5pt}]  (269.37,288.9) -- (299.32,288.9) ;
	\draw  [draw opacity=0][line width=1.5]  (249.75,307.44) .. controls (249.6,307.44) and (249.44,307.44) .. (249.29,307.44) .. controls (239.18,307.44) and (230.96,299.36) .. (230.74,289.3) -- (249.29,288.89) -- cycle ; \draw  [color={rgb, 255:red, 0; green, 119; blue, 255 }  ,draw opacity=1 ][line width=1.5]  (249.75,307.44) .. controls (249.6,307.44) and (249.44,307.44) .. (249.29,307.44) .. controls (239.18,307.44) and (230.96,299.36) .. (230.74,289.3) ;  
	\draw [color={rgb, 255:red, 0; green, 117; blue, 255 }  ,draw opacity=1 ][line width=1.5]    (248.87,270.35) .. controls (250.54,272.02) and (250.54,273.68) .. (248.87,275.35) .. controls (247.2,277.02) and (247.2,278.68) .. (248.87,280.35) .. controls (250.54,282.02) and (250.54,283.68) .. (248.87,285.35) .. controls (247.2,287.02) and (247.2,288.68) .. (248.87,290.35) .. controls (250.54,292.02) and (250.54,293.68) .. (248.87,295.35) .. controls (247.2,297.02) and (247.2,298.68) .. (248.87,300.35) .. controls (250.54,302.02) and (250.54,303.68) .. (248.87,305.35) -- (248.87,307.45) -- (248.87,307.45) ;
	\draw  [draw opacity=0][line width=0.75]  (230.74,288.71) .. controls (230.84,278.55) and (239.1,270.34) .. (249.29,270.34) .. controls (259.53,270.34) and (267.84,278.65) .. (267.84,288.89) .. controls (267.84,299.13) and (259.56,307.42) .. (249.33,307.44) -- (249.29,288.89) -- cycle ; \draw  [line width=0.75]  (230.74,288.71) .. controls (230.84,278.55) and (239.1,270.34) .. (249.29,270.34) .. controls (259.53,270.34) and (267.84,278.65) .. (267.84,288.89) .. controls (267.84,299.13) and (259.56,307.42) .. (249.33,307.44) ;  
	\draw  [fill={rgb, 255:red, 0; green, 0; blue, 0 }  ,fill opacity=1 ] (247.29,307.45) .. controls (247.29,306.58) and (248,305.88) .. (248.87,305.88) .. controls (249.74,305.88) and (250.44,306.58) .. (250.44,307.45) .. controls (250.44,308.32) and (249.74,309.02) .. (248.87,309.02) .. controls (248,309.02) and (247.29,308.32) .. (247.29,307.45) -- cycle ;
	\draw  [fill={rgb, 255:red, 0; green, 0; blue, 0 }  ,fill opacity=1 ] (247.29,270.35) .. controls (247.29,269.48) and (248,268.78) .. (248.87,268.78) .. controls (249.74,268.78) and (250.44,269.48) .. (250.44,270.35) .. controls (250.44,271.22) and (249.74,271.92) .. (248.87,271.92) .. controls (248,271.92) and (247.29,271.22) .. (247.29,270.35) -- cycle ;
	\draw    (139.33,288.44) -- (159.33,288.47) ;
	\draw    (139,232.77) -- (159,232.8) ;
	\draw    (138.67,188.1) -- (158.67,188.13) ;
	\draw    (139,93.44) -- (159,93.47) ;
	\draw  [fill={rgb, 255:red, 0; green, 0; blue, 0 }  ,fill opacity=1 ] (228.74,288.9) .. controls (228.74,288.03) and (229.45,287.33) .. (230.32,287.33) .. controls (231.19,287.33) and (231.89,288.03) .. (231.89,288.9) .. controls (231.89,289.77) and (231.19,290.47) .. (230.32,290.47) .. controls (229.45,290.47) and (228.74,289.77) .. (228.74,288.9) -- cycle ;
	\draw  [fill={rgb, 255:red, 0; green, 0; blue, 0 }  ,fill opacity=1 ] (266.22,288.9) .. controls (266.22,288.03) and (266.92,287.33) .. (267.79,287.33) .. controls (268.66,287.33) and (269.37,288.03) .. (269.37,288.9) .. controls (269.37,289.77) and (268.66,290.47) .. (267.79,290.47) .. controls (266.92,290.47) and (266.22,289.77) .. (266.22,288.9) -- cycle ;
	\draw  [dash pattern={on 4.5pt off 4.5pt}]  (200.6,446.69) -- (300.2,446.69) ;
	\draw    (241.26,397.09) .. controls (240.39,399.36) and (238.85,400.02) .. (236.63,399.07) .. controls (234.42,398.12) and (232.9,398.79) .. (232.07,401.07) .. controls (231.98,403.26) and (230.84,404.45) .. (228.65,404.62) .. controls (226.37,405.45) and (225.69,407) .. (226.62,409.27) .. controls (227.78,411.24) and (227.36,412.85) .. (225.37,414.1) .. controls (223.47,415.51) and (223.28,417.15) .. (224.81,419.03) .. controls (226.52,420.52) and (226.7,422.18) .. (225.34,424) .. controls (224.24,426.15) and (224.79,427.71) .. (226.98,428.69) .. controls (229.18,429.32) and (230,430.77) .. (229.45,433.03) .. controls (229.15,435.41) and (230.21,436.71) .. (232.63,436.94) .. controls (234.94,436.86) and (236.2,437.95) .. (236.39,440.2) .. controls (237.06,442.52) and (238.53,443.26) .. (240.8,442.42) .. controls (242.81,441.29) and (244.42,441.7) .. (245.62,443.63) .. controls (247.19,445.46) and (248.87,445.55) .. (250.66,443.9) .. controls (252.01,442.06) and (253.65,441.78) .. (255.56,443.05) .. controls (257.71,444.12) and (259.3,443.54) .. (260.31,441.32) .. controls (260.92,439.13) and (262.35,438.29) .. (264.59,438.82) .. controls (266.98,439.01) and (268.2,437.86) .. (268.25,435.38) .. controls (267.92,433.12) and (268.9,431.81) .. (271.19,431.44) .. controls (273.51,430.73) and (274.26,429.25) .. (273.45,426.98) .. controls (272.34,424.99) and (272.75,423.35) .. (274.68,422.07) .. controls (276.47,420.47) and (276.53,418.8) .. (274.84,417.07) .. controls (273.04,415.69) and (272.79,414.06) .. (274.09,412.17) .. controls (275.06,409.96) and (274.37,408.45) .. (272.03,407.64) .. controls (269.78,407.43) and (268.69,406.19) .. (268.76,403.92) .. controls (268.39,401.55) and (267.02,400.6) .. (264.63,401.09) .. controls (262.12,401.32) and (260.97,400.17) .. (261.18,397.64) -- (259.81,397.12) ;
	\draw  [color={rgb, 255:red, 0; green, 118; blue, 255 }  ,draw opacity=1 ][line width=1.5]  (241.26,397.09) .. controls (241.26,391.96) and (245.42,387.8) .. (250.55,387.8) .. controls (255.67,387.8) and (259.83,391.96) .. (259.83,397.09) .. controls (259.83,402.22) and (255.67,406.38) .. (250.55,406.38) .. controls (245.42,406.38) and (241.26,402.22) .. (241.26,397.09) -- cycle ;
	\draw  [fill={rgb, 255:red, 0; green, 0; blue, 0 }  ,fill opacity=1 ] (258.26,397.09) .. controls (258.26,396.22) and (258.96,395.52) .. (259.83,395.52) .. controls (260.7,395.52) and (261.41,396.22) .. (261.41,397.09) .. controls (261.41,397.96) and (260.7,398.67) .. (259.83,398.67) .. controls (258.96,398.67) and (258.26,397.96) .. (258.26,397.09) -- cycle ;
	\draw  [fill={rgb, 255:red, 0; green, 0; blue, 0 }  ,fill opacity=1 ] (239.68,397.09) .. controls (239.68,396.22) and (240.39,395.52) .. (241.26,395.52) .. controls (242.13,395.52) and (242.83,396.22) .. (242.83,397.09) .. controls (242.83,397.96) and (242.13,398.67) .. (241.26,398.67) .. controls (240.39,398.67) and (239.68,397.96) .. (239.68,397.09) -- cycle ;
	\draw    (137.97,446.7) -- (157.97,446.73) ;
	\draw  [fill={rgb, 255:red, 0; green, 0; blue, 0 }  ,fill opacity=1 ] (247.25,445.69) .. controls (247.25,444.82) and (247.96,444.12) .. (248.83,444.12) .. controls (249.7,444.12) and (250.4,444.82) .. (250.4,445.69) .. controls (250.4,446.56) and (249.7,447.27) .. (248.83,447.27) .. controls (247.96,447.27) and (247.25,446.56) .. (247.25,445.69) -- cycle ;
	\draw  [fill={rgb, 255:red, 208; green, 2; blue, 27 }  ,fill opacity=1 ] (248.33,188.6) .. controls (248.33,187.28) and (249.4,186.22) .. (250.71,186.22) .. controls (252.02,186.22) and (253.09,187.28) .. (253.09,188.6) .. controls (253.09,189.91) and (252.02,190.97) .. (250.71,190.97) .. controls (249.4,190.97) and (248.33,189.91) .. (248.33,188.6) -- cycle ;
	\draw  [draw opacity=0][line width=0.75]  (230.66,232.79) .. controls (231.04,222.87) and (239.19,214.95) .. (249.2,214.95) .. controls (259.04,214.95) and (267.09,222.61) .. (267.71,232.3) -- (249.2,233.5) -- cycle ; \draw  [color={rgb, 255:red, 0; green, 0; blue, 0 }  ,draw opacity=1 ][line width=0.75]  (230.66,232.79) .. controls (231.04,222.87) and (239.19,214.95) .. (249.2,214.95) .. controls (259.04,214.95) and (267.09,222.61) .. (267.71,232.3) ;  
	
	\draw (118.67,31.73) node [anchor=north west][inner sep=0.75pt]    {$M\ ( GeV)$};
	\draw (105.07,341.07) node [anchor=north west][inner sep=0.75pt]    {$10^{4}$};
	\draw (105.1,278.4) node [anchor=north west][inner sep=0.75pt]    {$10^{5}$};
	\draw (105.1,224.4) node [anchor=north west][inner sep=0.75pt]    {$10^{7}$};
	\draw (105.1,179.73) node [anchor=north west][inner sep=0.75pt]    {$10^{11}$};
	\draw (105.1,83.07) node [anchor=north west][inner sep=0.75pt]    {$10^{14}$};
	\draw (309,340.8) node [anchor=north west][inner sep=0.75pt]   [align=left] {KSVZ fermions~\cite{Farina:2013mla}};
	\draw (309,279.73) node [anchor=north west][inner sep=0.75pt]   [align=left] {vector leptoquark~\cite{FileviezPerez:2023rxn}};
	\draw (309,223.8) node [anchor=north west][inner sep=0.75pt]   [align=left] {type I see-saw~\cite{Vissani:1997ys}};
	\draw (309,179.1) node [anchor=north west][inner sep=0.75pt]  [color={rgb, 255:red, 208; green, 2; blue, 27 }  ,opacity=1 ] [align=left] {non-minimal coupling};
	\draw (309,84.4) node [anchor=north west][inner sep=0.75pt]   [align=left] {minimal coupling~\cite{deGouvea:2014xba}};
	\draw (274.2,167.39) node [anchor=north west][inner sep=0.75pt]  [font=\footnotesize,color={rgb, 255:red, 208; green, 2; blue, 27 }  ,opacity=1 ]  {$\xi $};
	\draw (105.1,438.13) node [anchor=north west][inner sep=0.75pt]    {$10^{3}$};
	\draw (309,437.7) node [anchor=north west][inner sep=0.75pt]   [align=left] {minimal dark matter~\cite{Cirelli:2005uq}};

\end{tikzpicture}

	\caption{\it Finite naturalness limits for new physics scenarios. Thick blue lines show heavy states with mass ${M}$ coupled to Standard Model particles (thin black lines in loops).}\label{fig:Schema}
\end{figure}
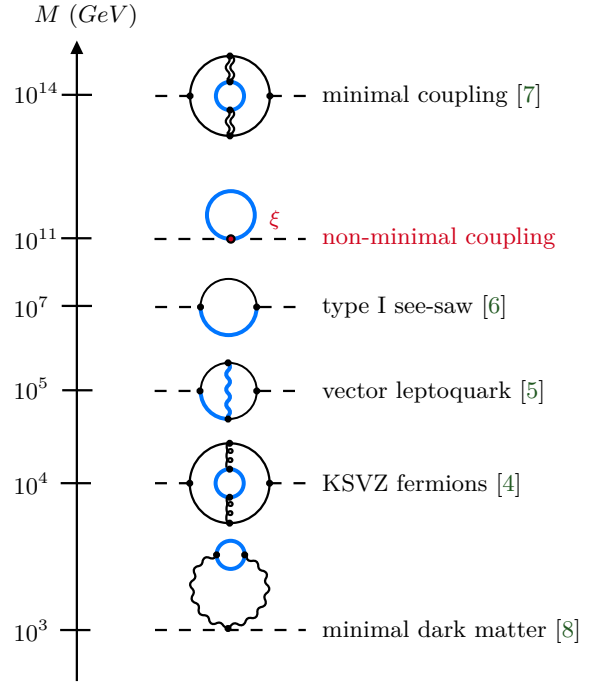

A practical, bottom-up possibility for reconciling the naturalness principle with experimental constraints is provided by the criterion of \textit{finite naturalness}, introduced in Ref.~\cite{Farina:2013mla} (see also~\cite{Giudice:2013yca}). The basic idea is that we should ignore quadratic divergences in loop computations. From the UV perspective, such an assumption may be justified  in models where an infinite tower of states at the new-physics scale cancels power divergences~\cite{Dienes:2001se, Volovik:2005zu}, or where the Planck scale arises from spontaneous symmetry breaking~\cite{Farina:2013mla}.
The finite naturalness principle guarantees that the Higgs mass remains small if there are no new heavy mass particles sizably coupled to the Higgs boson. Indeed, the Standard Model alone satisfies the criterion of finite naturalness, without demanding new physics up to the Planck scale~\cite{Farina:2013mla}. However, in extensions of the Standard Model that introduce particles with masses far above the weak scale, the Higgs mass generally acquires a quadratic dependence on these heavy scales—independent of the chosen regularization or subtraction scheme. 
Fig.\,\ref{fig:Schema} schematically illustrates the finite naturalness bound in relation to the presence of a mass scale $M$, directly or loop-coupled to the Higgs, as motivated by beyond-the-Standard-Model scenarios.

Hereafter, we define the Planck mass through $G = M_{\textrm{Pl}}^{-2}$, and the reduced Planck mass as $\bar{M}_{\textrm{Pl}} = M_{\textrm{Pl}} / \sqrt{8 \pi}$.

\vspace{0.1cm}
\textit{\textbf{The SM with a decoupled heavy scale.}} 
As a proxy for the existence of a heavy mass scale, we introduce a Dirac fermion $\Psi$ with mass $M_{\Psi}$ and without direct coupling to the Higgs boson. Therefore, contributions to the Higgs running mass $m_H$ proceed through the exchange of virtual gravitons. The common lore is that one needs (at least) three loops to produce an additive contribution to the Higgs running mass involving the new heavy scale~\cite{Farina:2013mla, deGouvea:2014xba}, $\delta m_h^2 \sim y_t^2 M_{\Psi}^6 / \bar{M}_{\textrm{Pl}}^4 (4\pi)^6$, resulting in a finite naturalness bound~\cite{deGouvea:2014xba} $ M_{\Psi} \lesssim 10^{14} \, {\rm GeV}$.

This result is derived assuming that the Higgs boson is minimally coupled to gravity. However, quantum fluctuations will unavoidably generate a dimension-four operator $\xi H^\dagger H R$, even if set to zero at  some scale~\cite{Espinosa:2015qea, Buchbinder:1992rb}.  Thus, the principle of renormalizability demands the addition of this operator to the original action from the beginning, inducing a non-minimal coupling to gravity.

We start with the Jordan frame action 
\begin{align}\label{eq:SJFSSBtwo}
\mathcal{S} &= 
\int d^4 x\sqrt{-g}
\bigg\{
\frac{1}{2}\left[
\bar{M}_{\textrm{Pl}}^2 + \xi(h+v)^2
\right]R \nonumber\\
&+ \frac{1}{2}(\partial_{\mu}h)(\partial^{\mu}h) - 
\left(
\frac{1}{2}m^2h^2 + \lambda v h^3 + \frac{\lambda}{4}h^4
\right)
\bigg\} \nonumber\\
& + S_m(\Psi, g_{\mu\nu}; M_\Psi)\,,
\,,
\end{align}
which combines the Einstein-Hilbert term ($R$ is the Ricci scalar), the Higgs sector of the SM after spontaneous symmetry breaking and the matter action $S_m$ describing the Dirac fermion $\Psi$ (decoupled from the SM). The Higgs mass is $m^2 = 2 \lambda v^2$. We work in the unitary gauge and neglect the interaction terms in the covariant derivatives.

The non-minimal coupling gives rise to a kinetic mixing between the Higgs and the graviton. The propagation of these two degrees of freedom is decoupled by means of a conformal transformation
\begin{align}\label{eq:Omegatwo}
g_{\mu\nu}\to \Omega^{2} g_{\mu\nu}\,,~~~~~\textrm{with:}~~~
 \Omega^2\equiv 1 + \frac{\xi(h+v)^2}{\bar{M}_{\textrm{Pl}}^2}\,.
\end{align}
Together with suitable field redefinitions, Eq.~\eqref{eq:Omegatwo} brings the original action Eq.~\eqref{eq:SJFSSBtwo} into the Einstein frame, where the Higgs field becomes indeed minimally coupled to gravity, but the scalar potential and the matter action are transformed (see Supplementary
Material for details).
%
%
%
The key aspect is that the fermion mass term becomes 
\begin{align}
-\frac{M_{\Psi}}{\Omega}\bar{\Psi}\Psi\,,
\end{align}
with $\Omega$ given in Eq.\,(\ref{eq:Omegatwo}). In the Einstein frame, therefore, the $\Psi$ mass generates a tower of interaction terms that couples $\Psi$ to the Higgs field. The leading correction to the Higgs mass in the limit $\xi \ll 1$ comes from the following effective operator
\begin{align}\label{eq:effective_operator}
\frac{\xi M_{\Psi}}{2\bar{M}_{\textrm{Pl}}^2}h^2\bar{\Psi}\Psi\,.
\end{align}
In Supplementary Material we also derive the same result working directly in the Jordan Frame, showing consistency among various approaches.

At one loop, the finite  contribution to the Higgs mass from the vertex Eq.~\eqref{eq:effective_operator} is
\begin{align}\label{eq:Higgs_mass_additive_term}
\delta m_h^2 = \xi \frac{4 M_{\Psi}^4}{(4\pi)^2 \bar{M}_{\textrm{Pl}}^2}\,,
\end{align}
which gives the following bound of naturalness on the mass-scale $M_{\Psi}$
\begin{align}
M_{\Psi} \lesssim  10^{10}\,\xi^{-1/4}\,\textrm{GeV}\,.\label{eq:NewNatBound}
\end{align}

This is more stringent than the bound given in Refs.~\cite{Farina:2013mla, deGouvea:2014xba}. Of course, Eq.\,(\ref{eq:NewNatBound}) depends on the non-minimal coupling $\xi$; however, it should be noted that the latter only enters with the scaling $\xi^{-1/4}$ which makes the dependence mild. Thus, even when $\xi$ is loop generated, i.e. $\xi \sim 1/ (4\pi)^2$, the corresponding bound is  $M_{\Psi}  \lesssim  10^{11}$.

The bound in Eq.~\eqref{eq:NewNatBound} holds also if we consider a scalar field instead of a Dirac fermion in the original action Eq.~\eqref{eq:SJFSSBtwo}, since the resulting mass term in going from the Jordan to the Einstein frame will again acquire a factor $\Omega^{-1}$.

\vspace{0.1cm}
\textit{\textbf{Application to inflationary theories.}} 
We consider an inflaton field $\phi$ with canonically normalized kinetic term and potential given, at the renormalizable level, by
\begin{align}
V(\phi) = \sum_{k=2}^{4}
\frac{a_k}{k!}\frac{g^{k-2}\phi^k}{M^{k-4}}\,,
\end{align}
where $g$ is some fundamental coupling, $M$ a mass scale and $a_k$ dimensionless coefficients.
%
We find it advantageous to keep track of the correct powers of coupling, along with those of mass, in the various terms of the potential. We further assume that the mass $M$ and the Planck scale are related by the equation $g\MPl=M$. The potential takes the form
\begin{align}
V(\phi) = M^2\MPl^2 
\sum_{k=2}^{4}\frac{a_k}{k!}\left(\frac{\phi}{\MPl}\right)^k\,.
\end{align}
We further define $c_k \equiv a_k/k!$, and 
$\bar{c}_k \equiv c_k/c_4$. We thus rewrite our potential in the form  
\begin{align}
V(x) = c_4M^2\MPl^2\left[
\bar{c}_2x^2 +
\bar{c}_3x^3 + x^4
\right]\,,\label{eq:FirstPote}
\end{align}
with $x\equiv \phi/\MPl$. The expression in square brackets is purely dimensionless, while the dimensionful part of the potential is completely captured by the factorized product  $M^2\MPl^2$. Field values are naturally given in units of the reduced Planck mass. 
The potential in Eq.\,(\ref{eq:FirstPote}) is nothing but a linear combination of monomial terms, each of which, if analyzed individually, is in tension with the upper limit on the tensor-to-scalar ratio~\cite{ACT:2025tim}. 
To overcome this issue, we consider a specific combination that yields the presence of an exact stationary inflection point at some field value $\phi_0$. 
In other words, we impose the conditions 
$V^{\prime}(\phi_0) = V^{\prime\prime}(\phi_0) = 0$. Solving the corresponding system of equations, we write
\begin{align}\label{eq:stationary_inflection_point_potential}
V(x) = c_4M^2\MPl^2
\left[
2x_0^2x^2
-
\frac{8x_0}{3}x^3 
+ 
x^4
\right]\,,
\end{align}
with  
$x_0\equiv \phi_0/\MPl$. 
In this form, the model features two free parameters. The overall scale of the potential, $V_0\equiv c_4 M^2\MPl^2$, and the parameter $x_0$ that controls the position of the stationary inflection point. In particular, only $x_0$ enters into the description of the inflationary dynamics, while $V_0$ is an overall normalization essentially determined by ensuring the match with the measured value of $A_s$. The model presented in Eq.~\eqref{eq:stationary_inflection_point_potential}, commonly referred to as inflection point inflation, has been discussed in several earlier works~\cite{Stewart:1996ey, Dimopoulos:2017xox,Allahverdi:2008bt,Baumann:2007ah,Baumann:2008kq,Krause:2007jk,Baumann:2007np,Drees:2021wgd}.
From the quadratic term in Eq.~\eqref{eq:stationary_inflection_point_potential} we identify the mass of the inflaton $m_{\phi}$ as
\begin{align}
m_{\phi}^2 \equiv 4x_0^2\left(
\frac{c_4M^2}{\MPl^2}
\right)\MPl^2\,.
\end{align}
We read the combination of parameters 
$c_4M^2/\MPl^2$ directly from the measured value of $A_s$ since in the slow-roll approximation we just have
\begin{align}
A_s = \frac{V(\phi_{*})}{24\pi^2\epsilon_V(\phi_{*}) \MPl^4}\,,\label{eq:SlowRollAs}
\end{align} 
where $\phi_\star$ is the field value at which cosmic microwave background (CMB) modes exited the Hubble horizon during
inflation, and $\epsilon_V$ is the slow-roll parameter (see Supplementary
Material for more details).

\begin{figure}[h]
	\centering
\includegraphics[width=0.495\textwidth]{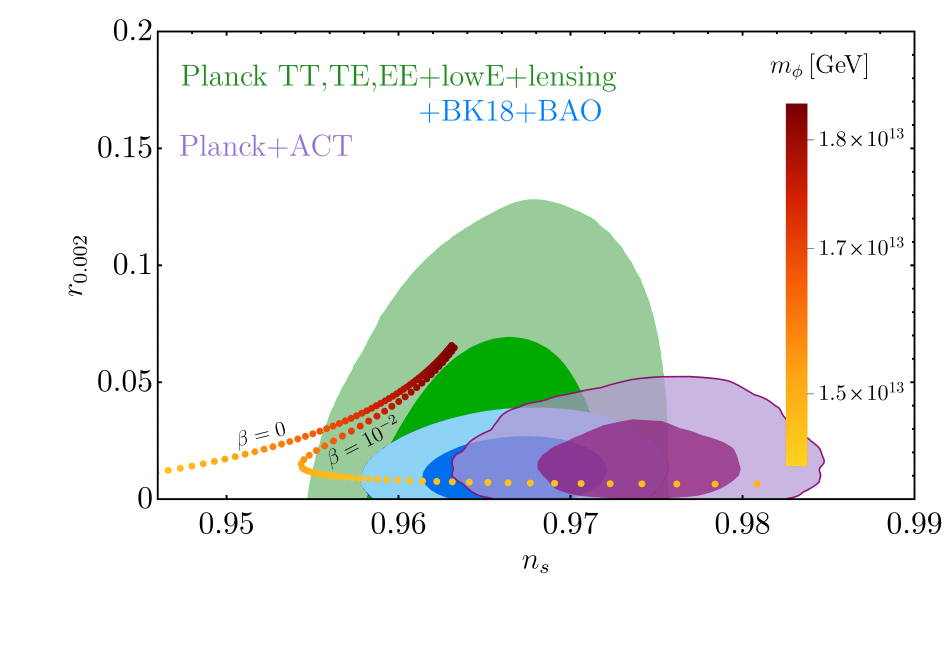}
	\caption{
    \it Prediction in the $(n_s,r)$ plane for the model given in 
Eq.~\eqref{eq:quasi_inflection_point}. We fix two characteristic values $\beta = 0, 10^{-2}$ while varying the parameter $x_0$. We keep track of the latter in the color bar, treating it as the mass of the inflaton $m_{\phi}$. We require $N=55$ $e$-folds of inflation after CMB modes exited the horizon. Experimental constraints are shown using the
Planck 2018 baseline analysis (green regions) and including BICEP/Keck and baryon acoustic oscillation
(BAO) data (blue regions), cf. Ref.\,\cite{BICEP:2021xfz}. 
We also show (purple regions) the recent results from the Atacama Cosmology Telescope (ACT), cf. Ref.\,\cite{ACT:2025tim}.
We show the 65\% and 95\% confidence contours. When the inflection point is exact ($\beta = 0$) the model is in tension with BK18+BAO and ACT data.}
\label{fig:CMBConstraint}
\end{figure}

Besides being in tension with experimental measurements (see Ref.~\cite{Drees:2021wgd} and Fig.\,\ref{fig:CMBConstraint}), an exact inflection point has a more severe problem.
The slow-roll trajectory identified by the solutions analyzed in Fig.\,\ref{fig:CMBConstraint} is not an attractor. The simplest way to convince oneself of this point is to calculate the number of e-folds. 
In slow-roll approximation, we find
\begin{align}
N& = -\int_{x_{*}}^{x_{\textrm{end}}}\frac{V(x)}{V^{\prime}(x)}dx  \nonumber\\
&= -\int_{x_{*}}^{x_{\textrm{end}}}
\frac{x(3x^2 - 8xx_0 +6x_0^2)}{12(x-x_0)^2}dx\,,\label{eq:EfoldNumberExact}
\end{align}
where $x_*=\phi_*/\MPl$ and $x_{\rm{end}}=\phi_{\rm{end}}/\MPl$. The integrand function in Eq.\,(\ref{eq:EfoldNumberExact}) features a singularity at $x=x_0$. Since, by construction, $x_{\textrm{end}} < x_0$, we are compelled to choose $x_{*} < x_0$ to avoid encountering the singularity. 
Indeed, in the analysis leading to Fig.\,\ref{fig:CMBConstraint}, 
we imposed the condition 
$0<x_{\textrm{end}}<x_* < x_0$. 
Clearly, there is nothing mathematically wrong in restricting our analysis to this choice.
\begin{figure}[h]
	\centering
\includegraphics[width=0.495\textwidth]{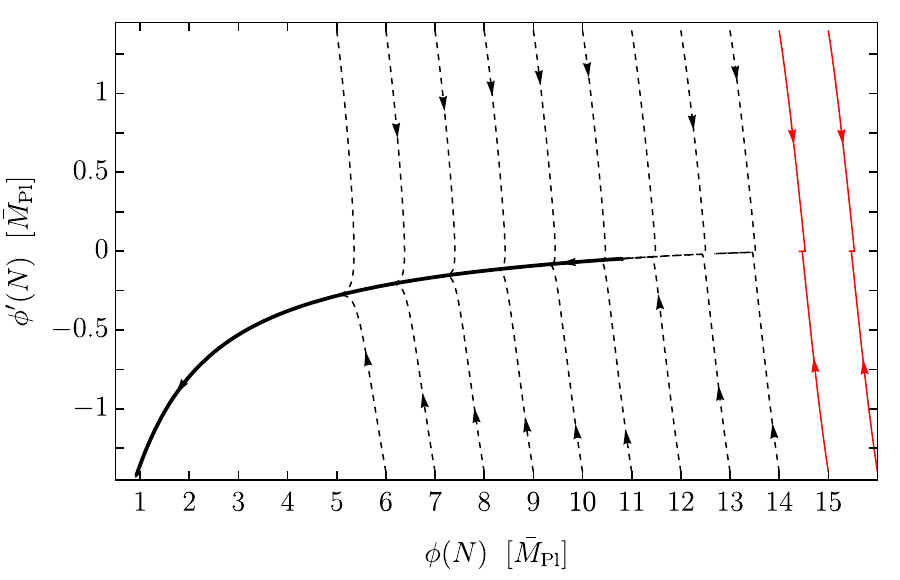}
	\caption{\it Phase-space portrait of the classical inflationary dynamics for $x_0 = 15$ ($\beta = 0$).  The slow-roll trajectory associated to observable inflation is indicated in solid bold black. The trajectories marked in red are disconnected from the rest of the phase space.}
\label{fig:PhaseSpacePortrait}
\end{figure}

We confirm the previous findings in slow-roll approximation by showing in Fig.\,\ref{fig:PhaseSpacePortrait} a phase-space portrait of the classical inflationary dynamics, obtained solving exactly Eq.\,(\ref{eq:EoM}) with arbitrary initial conditions. As a benchmark value, we consider $x_0 = 15$. 
The trajectories marked in red are disconnected from the rest of the phase space. 
Following these trajectories, the inflaton stops at around the inflection point, and inflation never ends.
On the contrary, all trajectories indicated with dashed black lines 
converge onto the slow-roll trajectory, indicated in solid bold black, rolling along which the inflaton first reaches the condition $\epsilon > 1$ where the accelerated expansion ends and subsequently the minimum at $x=0$ where reheating oscillations begin. 
 
In the next section, we show that is it possible to modify the model in such a way that slow-roll becomes an attractor of the dynamics. 

\vspace{0.1cm}
\textit{\textbf{Quasi-stationary inflection point inflation.}} 
We perturb the stationary inflection point~\cite{Drees:2021wgd} allowing the cubic term in Eq.~\eqref{eq:stationary_inflection_point_potential} to deviate by a factor $1-\beta$, i.e.
\begin{equation}\label{eq:quasi_inflection_point}
    V(x) = c_4M^2\MPl^2
\left[
2x_0^2x^2
-
\frac{8x_0}{3}(1-\beta)x^3 
+ 
x^4
\right]\,.
\end{equation}
\begin{figure}[!htb]
	\centering
\includegraphics[width=0.495\textwidth]{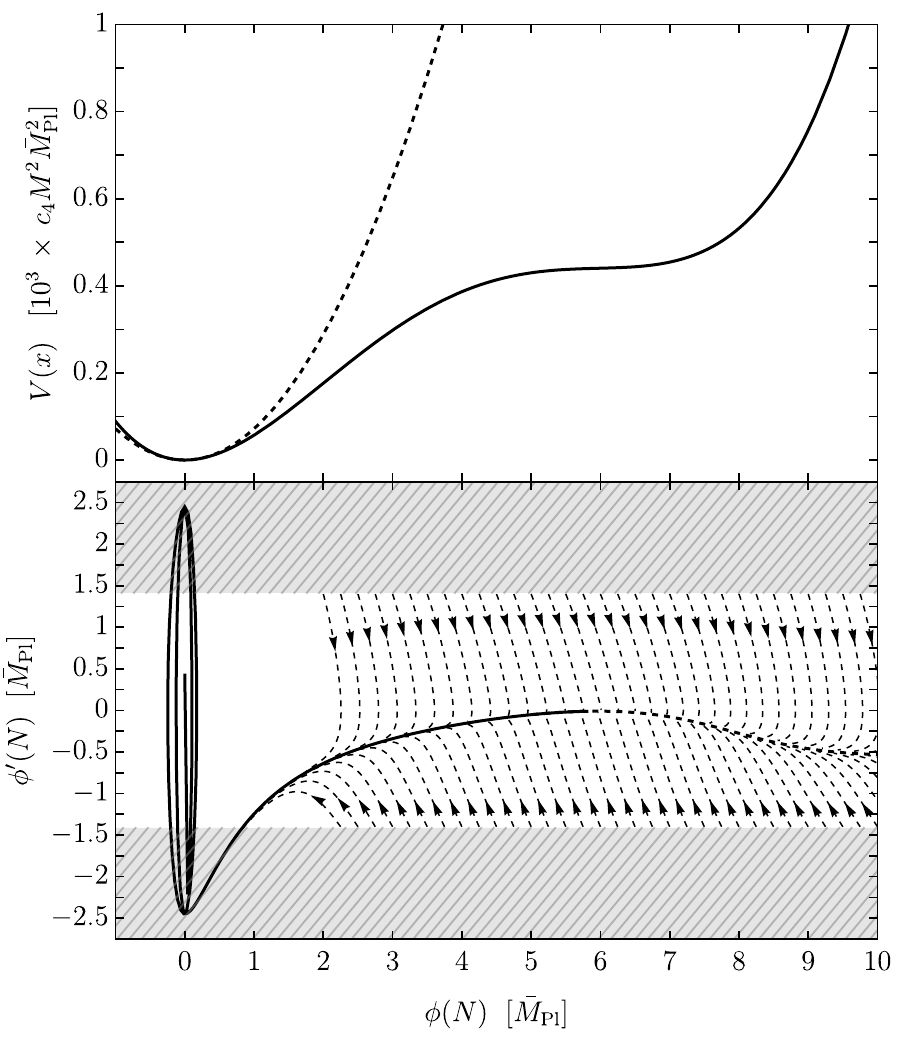}
	\caption{\it Phase-space portrait of the classical inflationary dynamics for $x_0 = 6$ and $\beta = 10^{-3}$. Varying $\beta$ does not qualitatively change the behaviors of the dynamics. The point along the slow-roll trajectory that marks the transition between the continuous and dashed part corresponds to $x_*$, i.e. the field value at which the CMB modes exited the horizon during inflation.
 }
\label{fig:BackgroundDyn}
\end{figure}
\begin{figure}[!htb]
	\centering
    \includegraphics[width=0.495\textwidth]{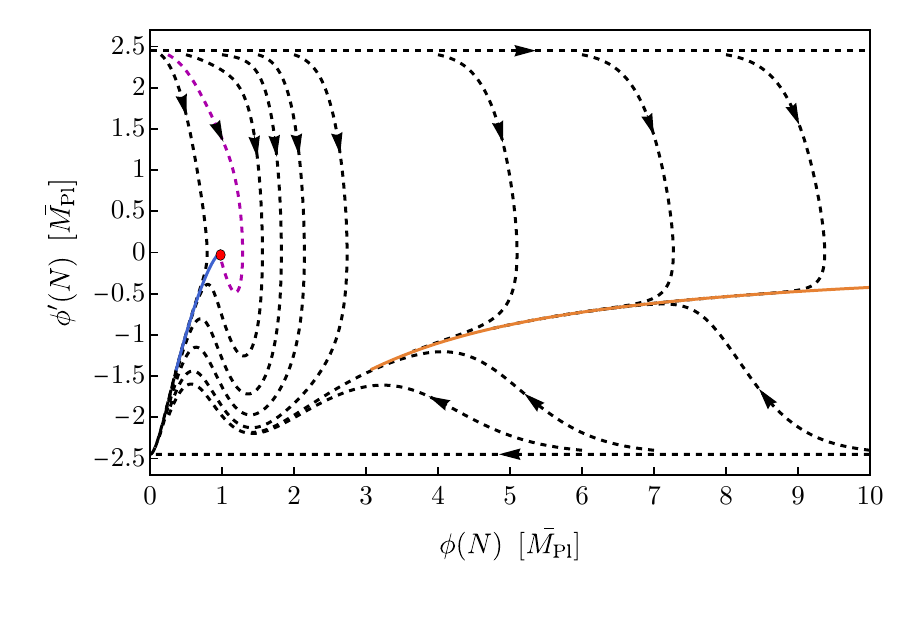}
	\caption{\it Phase-space portrait of the classical inflationary dynamics for $x_0 = 1$ and $\beta = 1.85 \times 10^{-6}$. Varying $\beta$ does not qualitatively change the behaviors of the dynamics. There are two
distinct trajectories that contains a phase of slow-roll
inflation, marked in blue and in orange. The red point along the blue trajectory corresponds to $x_* \approx 0.9992$, i.e. the field value at which the CMB modes exited the horizon during inflation. The resulting observables are $n_s = 0.965$ and $r = 2 \times 10^{-8}$. 
 }
\label{fig:BackgroundDyn2}
\end{figure}
With this expression, both $V'(x_0)$ and $V''(x_0)$ are proportional to $\beta$. Thus, assuming $\beta \ll 1$, $x_0$ becomes an approximate inflection point. This simple generalization introduces new features. First of all, the model with $\beta \neq 0 $ is able to reproduce the experimental data, as shown in Fig.\,\ref{fig:CMBConstraint}, where we plot the model's prediction in the $(n_s,r)$ plane for the two characteristic values of $\beta = 0, 10^{-2}$, varying the parameter $x_0$ (see Supplementary
Material for details about the procedure we use to extract cosmological observables) and requiring $N=55$ $e$-folds of inflation after CMB modes exited the horizon.

Moreover, in Fig.~\ref{fig:BackgroundDyn} we show the phase-space portrait of the classical inflationary dynamics using the benchmark value $x_0 = 6$. At the variance with the previous case, there are no more curves disconnected from the slow-roll trajectory, meaning that the latter becomes an attractor of the dynamics. Hence, we conclude that in the large-field regime ($x_0 > 1$) the inflationary dynamics is robust against perturbations of the initial condition. However, this conclusion has to be changed in the limit of small-field values ($x_0 \lesssim 1$). In Fig.~\ref{fig:BackgroundDyn2}, we show again the phase-space portrait of the inflationary dynamics, but fixing $x_0 = 1$. Remarkably, there are two distinct trajectories that contain a phase of slow-roll inflation. The first one, marked in blue, acts as an attractor for trajectories with initial position $\phi(N = 0) \lesssim 2 \,\bar{M}_{\rm Pl}$, whereas the second one, highlighted in orange, is an attractor when the initial position is $\phi(N = 0) \gtrsim 2 \,\bar{M}_{\rm Pl}$. In the latter case, the slow-roll parameters are kept small since the field goes through super-Planckian values and the quasi-inflection point in $x_0 = 1$ does not play any role. Indeed, the orange curve in Fig.~\ref{fig:BackgroundDyn2} is qualitatively the same slow-roll phase generated by a generic combination of monomials, incompatible with the Planck measurements. Vice versa, the slow-roll parameters associated to the blue curve in Fig.~\ref{fig:BackgroundDyn2} are small thanks to the quasi-stationary inflection point. 
In this case, $N \approx 55$ $e$-folds of observed inflation require the CMB modes to exit the horizon at $x_* \approx 0.9992$ (see the red point in Fig.~\ref{fig:BackgroundDyn2}), i.e. slightly smaller than $x_0$. However, the latter point is not generically reached for trajectories with initial position $\phi(N = 0) \lesssim 2$ and thus the model requires fine-tuning. To clarify this property of the model, let us consider the dashed magenta curve depicted in Fig.~\ref{fig:BackgroundDyn2}, which goes through $x_*$ and gives rise to inflationary observables compatible with Planck data. We see from Fig.~\ref{fig:BackgroundDyn2} that a small perturbation on this trajectory generates a new curve that does not go through $x_*$ (even if it will generically fall onto the slow-roll trajectory in blue). These results highlight that the small-field regime is not robust against perturbation on the initial conditions, even if capable of reproducing experimental data. 


The presence of two distinct slow-roll trajectories in the small-field regime is better understood by studying the solutions to the equation $\epsilon_V = 1$, where $\epsilon_V$ is 
defined in Eq.~\eqref{eq:slow_roll_parameters}. It turns out that for $x_0 \gg 1$ there is only one positive solution, confirming the existence of just one slow-roll trajectory. On the other hand, for $x_0 \lesssim 1$ different solutions arise, allowing for the existence of a second slow-roll trajectory. In particular, it is possible to analytically compute the point where the equation $\epsilon_V = 1$ develops multiple positive solutions for $\beta \ll 1$ as $x_0 \approx 2.3$.



\begin{figure}[!htb] 
	\centering
	\includegraphics[width=0.495\textwidth]{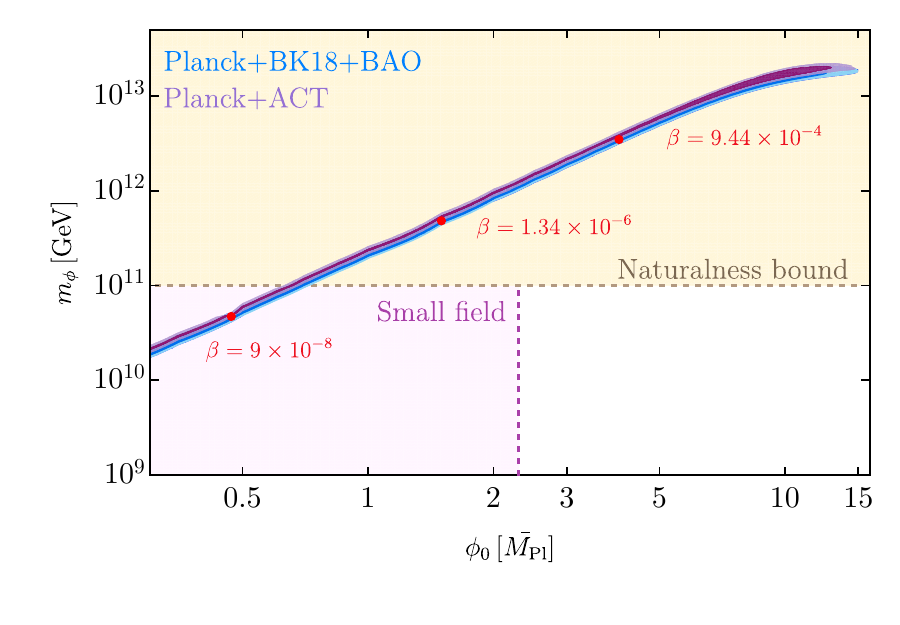}
	\caption{\it Parameter space of the model considered in Eq.~\eqref{eq:quasi_inflection_point}, shown for different values of the $\beta$ parameter. We fix $\xi = 1/(4\pi)^2$, assuming it is loop-generated. The resulting excluded region, corresponding to $m_{\phi} \gtrsim 10^{11} , \mathrm{GeV}$, is shown in beige. The small-field regime, $x_0 \lesssim 2.3$, is indicated in light pink. The region of parameter space compatible with Planck data (blue/azure region), for $N \approx 55$ $e$-folds of inflation, either lies within the region excluded by finite naturalness or falls into the small-field regime, where inflationary dynamics requires significant fine-tuning (see Fig.~\ref{fig:BackgroundDyn2}). The ACT data (purple region) further accentuate this trend. 
		As highlighted in the discussion, the naturalness bound depends only mildly on $\xi$.
	}
	\label{fig:InflationFit}
\end{figure}

\vspace{0.1cm}
\textit{\textbf{How robust is single-field inflation?}} 
%
In Fig.~\ref{fig:InflationFit} we show the allowed parameter space of the model, together with different values of the inflaton mass. In the large-field regime ($x_0 \gtrsim 1$), as pointed out in the previous section, inflation is robust against perturbations of the initial conditions. However, Fig.~\ref{fig:InflationFit} shows that the corresponding values of inflaton mass violate the naturalness bound Eq.~\eqref{eq:NewNatBound}, resulting in a destabilization of the running Higgs mass. To avoid this problem, we are forced to work in the small-field regime ($x_0 \lesssim 1$), where $m_\phi \lesssim 10^{11} \, {\rm GeV}$. Nevertheless, we already highlighted that in this regime a successful inflationary dynamics requires a considerable amount of fine-tuning. Furthermore, the recent results from ACT~\cite{ACT:2025tim} pushes the viable region in the parameter space even further into the disfavored regime, as visible in Fig.~\ref{fig:InflationFit}.
We emphasize that introducing a direct coupling between the inflaton and the Standard Model, such as $\lambda \phi^2 h^2$, worsens this issue, as the bound in Eq.~\eqref{eq:NewNatBound} would become even more stringent.

\vspace{0.1cm}
\textit{\textbf{Conclusions.}} In this work, we showed that in the limiting case of an inflationary sector $\mathcal{L}_\phi$ that couples only gravitationally to the SM, avoiding fine-tuning is not an option. In particular, in the large field regime inflationary dynamics provides a compelling dynamical solution to the origin of the observed universe. However, the required inflaton mass (assuming the finite naturalness criterion) is in conflict with the weak scale, even if direct couplings to the Standard Model are absent. This is because the principle of renormalizability demands a non-minimal coupling of the Higgs to gravity, resulting in a contribution to the running Higgs mass well before the three-loop diagrams previously discussed in the literature~\cite{deGouvea:2014xba}. On the other hand, a light inflation would pacifically coexist with the weak scale, but implies a considerable amount of fine-tuning on the inflationary dynamics, as also shown in Ref.~\cite{Clough:2016ymm}. 

Although we adopted a specific inflationary model to streamline our argument, our main claim ultimately relies on only two general assumptions:
\begin{itemize}[itemsep=0.1em, topsep=0.3em]
\item small-field inflation (i.e., when the field value $\phi$ is smaller than $\bar{M}_{\rm Pl}$) requires significant fine-tuning of the inflaton initial conditions;
\item the typical mass scale of inflationary models operating in the large-field regime is $m_\phi \sim 10^{13}\,{\rm GeV}$.
\end{itemize}
To the best of our knowledge, these are generic features of single-field inflationary models, most notably Starobinsky inflation~\cite{Ketov:2025nkr}.

There are numerous avenues for future research. 
First of all, we note that although $\phi(N)$ and $\phi’(N)$ define an effective phase space, they do not constitute canonical variables~\cite{Remmen:2013eja}. Consequently, a rigorous quantification of the fine-tuning in the small-field regime requires introducing a suitable canonical measure (see e.g. Ref.~\cite{Remmen:2014mia}).
Additionally, our findings might be used as a theoretical guide for classes of models that elegantly evade the naturalness bound while still operating in the large-field regime. In this regard, Higgs inflation is not admissible~\cite{Barbon:2009ya,Masina:2018ejw}.
On the other hand, embedding the inflaton field within a supersymmetric sector may provide a viable approach.
At the same time, our work motivates the exploration of alternative cosmological scenarios, such as bouncing cosmologies~\cite{Battefeld:2014uga, Brandenberger:2016vhg, DelGrosso:2024gnj}
%
.
These questions, however, are beyond the scope of the present work and are left for future investigation.

\vspace{0.1cm}
\textit{\textbf{Acknowledgments.}}
We thank David E. Kaplan, Francesco Serra, and Roberto Contino for interesting discussions. We thank Grant Remmen for useful comments on the manuscript. L.D.G. is supported by NSF Grants No. PHY-2207502, AST-2307146, PHY-090003
and PHY-20043, by NASA Grant No. 21-ATP21-0010,
by the John Templeton Foundation Grant 62840, and by
the Simons Investigator Grant No. 144924. 
This work is partially supported by ICSC - Centro
Nazionale di Ricerca in High Performance Computing,
Big Data and Quantum Computing, funded by European
Union-NextGenerationEU and by the research grant number 20227S3M3B
under the program PRIN 2022 of
the Italian Ministero dell’Università e Ricerca (MUR). 

\bibliography{main}

@article{Branchina:2022gll,
	author = "Branchina, Carlo and Branchina, Vincenzo and Contino, Filippo",
	title = "{Physical tuning and naturalness}",
	eprint = "2208.05431",
	archivePrefix = "arXiv",
	primaryClass = "hep-ph",
	doi = "10.1103/PhysRevD.107.096012",
	journal = "Phys. Rev. D",
	volume = "107",
	number = "9",
	pages = "096012",
	year = "2023"
}

@article{Remmen:2014mia,
    author = "Remmen, Grant N. and Carroll, Sean M.",
    title = "{How Many $e$-Folds Should We Expect from High-Scale Inflation?}",
    eprint = "1405.5538",
    archivePrefix = "arXiv",
    primaryClass = "hep-th",
    reportNumber = "CALT-2014-138",
    doi = "10.1103/PhysRevD.90.063517",
    journal = "Phys. Rev. D",
    volume = "90",
    number = "6",
    pages = "063517",
    year = "2014"
}

@article{Remmen:2013eja,
    author = "Remmen, Grant N. and Carroll, Sean M.",
    title = "{Attractor Solutions in Scalar-Field Cosmology}",
    eprint = "1309.2611",
    archivePrefix = "arXiv",
    primaryClass = "gr-qc",
    reportNumber = "CALT-68-2853",
    doi = "10.1103/PhysRevD.88.083518",
    journal = "Phys. Rev. D",
    volume = "88",
    pages = "083518",
    year = "2013"
}

@article{Volovik:2005zu,
    author = "Volovik, G. E.",
    title = "{Vacuum energy: Quantum hydrodynamics versus quantum gravity}",
    eprint = "gr-qc/0505104",
    archivePrefix = "arXiv",
    doi = "10.1134/1.2137368",
    journal = "JETP Lett.",
    volume = "82",
    pages = "319--324",
    year = "2005"
}

@article{ACT:2025tim,
    author = "Calabrese, Erminia and others",
    collaboration = "ACT",
    title = "{The Atacama Cosmology Telescope: DR6 Constraints on Extended Cosmological Models}",
    eprint = "2503.14454",
    archivePrefix = "arXiv",
    primaryClass = "astro-ph.CO",
    reportNumber = "FERMILAB-PUB-25-0157-PPD",
    month = "3",
    year = "2025",
journal = ""
}

@article{Masina:2018ejw,
    author = "Masina, Isabella",
    title = "{Ruling out Critical Higgs Inflation?}",
    eprint = "1805.02160",
    archivePrefix = "arXiv",
    primaryClass = "hep-ph",
    reportNumber = "CP3-ORIGINS-2018-016, CERN-TH-2018-111, CP3-Origins-2018-016, DNRF90",
    doi = "10.1103/PhysRevD.98.043536",
    journal = "Phys. Rev. D",
    volume = "98",
    number = "4",
    pages = "043536",
    year = "2018"
}

@article{Barbon:2009ya,
    author = "Barbon, J. L. F. and Espinosa, J. R.",
    title = "{On the Naturalness of Higgs Inflation}",
    eprint = "0903.0355",
    archivePrefix = "arXiv",
    primaryClass = "hep-ph",
    reportNumber = "IFT-UAM-CSIC-09-10, UAB-FT-665",
    doi = "10.1103/PhysRevD.79.081302",
    journal = "Phys. Rev. D",
    volume = "79",
    pages = "081302",
    year = "2009"
}

@inproceedings{Ketov:2025nkr,
    author = "Ketov, Sergei V.",
    title = "{On Legacy of Starobinsky Inflation}",
    eprint = "2501.06451",
    archivePrefix = "arXiv",
    primaryClass = "gr-qc",
    reportNumber = "IPMU25-0001",
    month = "1",
    year = "2025"
}

@article{Baumann:2007np,
    author = "Baumann, Daniel and Dymarsky, Anatoly and Klebanov, Igor R. and McAllister, Liam and Steinhardt, Paul J.",
    title = "{A Delicate universe}",
    eprint = "0705.3837",
    archivePrefix = "arXiv",
    primaryClass = "hep-th",
    reportNumber = "ITEP-TH-12-07, PUPT-2235",
    doi = "10.1103/PhysRevLett.99.141601",
    journal = "Phys. Rev. Lett.",
    volume = "99",
    pages = "141601",
    year = "2007"
}

@article{Krause:2007jk,
    author = "Krause, Axel and Pajer, Enrico",
    title = "{Chasing brane inflation in string-theory}",
    eprint = "0705.4682",
    archivePrefix = "arXiv",
    primaryClass = "hep-th",
    reportNumber = "LMU-ASC-35-07",
    doi = "10.1088/1475-7516/2008/07/023",
    journal = "JCAP",
    volume = "07",
    pages = "023",
    year = "2008"
}

@article{Baumann:2008kq,
    author = "Baumann, Daniel and Dymarsky, Anatoly and Kachru, Shamit and Klebanov, Igor R. and McAllister, Liam",
    title = "{Holographic Systematics of D-brane Inflation}",
    eprint = "0808.2811",
    archivePrefix = "arXiv",
    primaryClass = "hep-th",
    reportNumber = "SLAC-PUB-13365, SU-ITP-08-19, ITEP-TH-17-08",
    doi = "10.1088/1126-6708/2009/03/093",
    journal = "JHEP",
    volume = "03",
    pages = "093",
    year = "2009"
}

@article{Baumann:2007ah,
    author = "Baumann, Daniel and Dymarsky, Anatoly and Klebanov, Igor R. and McAllister, Liam",
    title = "{Towards an Explicit Model of D-brane Inflation}",
    eprint = "0706.0360",
    archivePrefix = "arXiv",
    primaryClass = "hep-th",
    reportNumber = "PUPT-2237, ITEP-TH-22-07",
    doi = "10.1088/1475-7516/2008/01/024",
    journal = "JCAP",
    volume = "01",
    pages = "024",
    year = "2008"
}

@article{Allahverdi:2008bt,
    author = "Allahverdi, Rouzbeh and Dutta, Bhaskar and Mazumdar, Anupam",
    title = "{Attraction towards an inflection point inflation}",
    eprint = "0806.4557",
    archivePrefix = "arXiv",
    primaryClass = "hep-ph",
    reportNumber = "MIFP-08-14",
    doi = "10.1103/PhysRevD.78.063507",
    journal = "Phys. Rev. D",
    volume = "78",
    pages = "063507",
    year = "2008"
}

@article{Dimopoulos:2017xox,
    author = "Dimopoulos, Konstantinos and Owen, Charlotte and Racioppi, Antonio",
    title = "{Loop inflection-point inflation}",
    eprint = "1706.09735",
    archivePrefix = "arXiv",
    primaryClass = "hep-ph",
    doi = "10.1016/j.astropartphys.2018.06.002",
    journal = "Astropart. Phys.",
    volume = "103",
    pages = "16--20",
    year = "2018"
}

@article{Stewart:1996ey,
    author = "Stewart, Ewan D.",
    title = "{Flattening the inflaton's potential with quantum corrections}",
    eprint = "hep-ph/9606241",
    archivePrefix = "arXiv",
    reportNumber = "RESCEU-19-96",
    doi = "10.1016/S0370-2693(96)01458-X",
    journal = "Phys. Lett. B",
    volume = "391",
    pages = "34--38",
    year = "1997"
}

@article{DelGrosso:2024gnj,
    author = "Del Grosso, Loris and Kaplan, David E. and Melia, Tom and Poulin, Vivian and Rajendran, Surjeet and Smith, Tristan L.",
    title = "{Cosmological Consequences of Unconstrained Gravity and Electromagnetism}",
    eprint = "2405.06374",
    archivePrefix = "arXiv",
    primaryClass = "hep-ph",
    reportNumber = "FERMILAB-PUB-24-0491-SQMS-V",
    month = "5",
    year = "2024",
journal = ""
}

@article{Giudice:2013yca,
    author = "Giudice, Gian F.",
    title = "{Naturalness after LHC8}",
    eprint = "1307.7879",
    archivePrefix = "arXiv",
    primaryClass = "hep-ph",
    reportNumber = "CERN-PH-TH-2013-180",
    doi = "10.22323/1.180.0163",
    journal = "PoS",
    volume = "EPS-HEP2013",
    pages = "163",
    year = "2013"
}

@article{Cirelli:2005uq,
    author = "Cirelli, Marco and Fornengo, Nicolao and Strumia, Alessandro",
    title = "{Minimal dark matter}",
    eprint = "hep-ph/0512090",
    archivePrefix = "arXiv",
    reportNumber = "DFTT40-2005, IFUP-TH-2005-34",
    doi = "10.1016/j.nuclphysb.2006.07.012",
    journal = "Nucl. Phys. B",
    volume = "753",
    pages = "178--194",
    year = "2006"
}

@article{FileviezPerez:2023rxn,
    author = "Fileviez P\'erez, Pavel and Murgui, Clara and Patrone, Samuel and Testa, Adriano and Wise, Mark B.",
    title = "{Finite naturalness and quark-lepton unification}",
    eprint = "2308.07367",
    archivePrefix = "arXiv",
    primaryClass = "hep-ph",
    reportNumber = "Report-no: CALT-TH/2023-025",
    doi = "10.1103/PhysRevD.109.015011",
    journal = "Phys. Rev. D",
    volume = "109",
    number = "1",
    pages = "015011",
    year = "2024"
}

@article{Vissani:1997ys,
    author = "Vissani, Francesco",
    title = "{Do experiments suggest a hierarchy problem?}",
    eprint = "hep-ph/9709409",
    archivePrefix = "arXiv",
    reportNumber = "IC-97-148",
    doi = "10.1103/PhysRevD.57.7027",
    journal = "Phys. Rev. D",
    volume = "57",
    pages = "7027--7030",
    year = "1998"
}

@article{RevModPhys.55.583,
  title = {The renormalization group and critical phenomena},
  author = {Wilson, Kenneth G.},
  journal = {Rev. Mod. Phys.},
  volume = {55},
  issue = {3},
  pages = {583--600},
  numpages = {0},
  year = {1983},
  month = {Jul},
  publisher = {American Physical Society},
  doi = {10.1103/RevModPhys.55.583},
  url = {https://link.aps.org/doi/10.1103/RevModPhys.55.583}
}

@article{Craig:2022eqo,
	author = "Craig, Nathaniel",
	title = "{Naturalness: past, present, and future}",
	eprint = "2205.05708",
	archivePrefix = "arXiv",
	primaryClass = "hep-ph",
	doi = "10.1140/epjc/s10052-023-11928-7",
	journal = "Eur. Phys. J. C",
	volume = "83",
	number = "9",
	pages = "825",
	year = "2023"
}

@article{Battefeld:2014uga,
	author = "Battefeld, D. and Peter, Patrick",
	title = "{A Critical Review of Classical Bouncing Cosmologies}",
	eprint = "1406.2790",
	archivePrefix = "arXiv",
	primaryClass = "astro-ph.CO",
	doi = "10.1016/j.physrep.2014.12.004",
	journal = "Phys. Rept.",
	volume = "571",
	pages = "1--66",
	year = "2015"
}

@article{Brandenberger:2016vhg,
	author = "Brandenberger, Robert and Peter, Patrick",
	title = "{Bouncing Cosmologies: Progress and Problems}",
	eprint = "1603.05834",
	archivePrefix = "arXiv",
	primaryClass = "hep-th",
	doi = "10.1007/s10701-016-0057-0",
	journal = "Found. Phys.",
	volume = "47",
	number = "6",
	pages = "797--850",
	year = "2017"
}

@book{Buchbinder:1992rb,
	author = "Buchbinder, I. L. and Odintsov, S. D. and Shapiro, I. L.",
	title = "{Effective action in quantum gravity}",
	year = "1992",
publisher = "CRC Press"
}

@article{Dienes:2001se,
	author = "Dienes, Keith R.",
	title = "{Solving the hierarchy problem without supersymmetry or extra dimensions: An Alternative approach}",
	eprint = "hep-ph/0104274",
	archivePrefix = "arXiv",
	doi = "10.1016/S0550-3213(01)00344-3",
	journal = "Nucl. Phys. B",
	volume = "611",
	pages = "146--178",
	year = "2001"
}

@article{Farina:2013mla,
	author = "Farina, Marco and Pappadopulo, Duccio and Strumia, Alessandro",
	title = "{A modified naturalness principle and its experimental tests}",
	eprint = "1303.7244",
	archivePrefix = "arXiv",
	primaryClass = "hep-ph",
	doi = "10.1007/JHEP08(2013)022",
	journal = "JHEP",
	volume = "08",
	pages = "022",
	year = "2013"
}

@article{BICEP:2021xfz,
    author = "Ade, P. A. R. and others",
    collaboration = "BICEP, Keck",
    title = "{Improved Constraints on Primordial Gravitational Waves using Planck, WMAP, and BICEP/Keck Observations through the 2018 Observing Season}",
    eprint = "2110.00483",
    archivePrefix = "arXiv",
    primaryClass = "astro-ph.CO",
    doi = "10.1103/PhysRevLett.127.151301",
    journal = "Phys. Rev. Lett.",
    volume = "127",
    number = "15",
    pages = "151301",
    year = "2021"
}

@article{Drees:2021wgd,
    author = "Drees, Manuel and Xu, Yong",
    title = "{Small field polynomial inflation: reheating, radiative stability and lower bound}",
    eprint = "2104.03977",
    archivePrefix = "arXiv",
    primaryClass = "hep-ph",
    doi = "10.1088/1475-7516/2021/09/012",
    journal = "JCAP",
    volume = "09",
    pages = "012",
    year = "2021"
}

@article{Clough:2016ymm,
    author = "Clough, Katy and Lim, Eugene A. and DiNunno, Brandon S. and Fischler, Willy and Flauger, Raphael and Paban, Sonia",
    title = "{Robustness of Inflation to Inhomogeneous Initial Conditions}",
    eprint = "1608.04408",
    archivePrefix = "arXiv",
    primaryClass = "hep-th",
    reportNumber = "KCL-PH-TH-2016-53",
    doi = "10.1088/1475-7516/2017/09/025",
    journal = "JCAP",
    volume = "09",
    pages = "025",
    year = "2017"
}

@article{Espinosa:2015qea,
    author = "Espinosa, Jose R. and Giudice, Gian F. and Morgante, Enrico and Riotto, Antonio and Senatore, Leonardo and Strumia, Alessandro and Tetradis, Nikolaos",
    title = "{The cosmological Higgstory of the vacuum instability}",
    eprint = "1505.04825",
    archivePrefix = "arXiv",
    primaryClass = "hep-ph",
    reportNumber = "CERN-PH-TH-2015-119",
    doi = "10.1007/JHEP09(2015)174",
    journal = "JHEP",
    volume = "09",
    pages = "174",
    year = "2015"
}

@article{deGouvea:2014xba,
    author = "de Gouvea, Andre and Hernandez, Daniel and Tait, Tim M. P.",
    title = "{Criteria for Natural Hierarchies}",
    eprint = "1402.2658",
    archivePrefix = "arXiv",
    primaryClass = "hep-ph",
    reportNumber = "NUHEP-TH-14-01, UCI-HEP-TR-2013-20",
    doi = "10.1103/PhysRevD.89.115005",
    journal = "Phys. Rev. D",
    volume = "89",
    number = "11",
    pages = "115005",
    year = "2014"
}

\clearpage
\appendix
\renewcommand{\theequation}{S\arabic{equation}}
\section{Supplemental Material}
\setcounter{secnumdepth}{2}
\setcounter{equation}{0}
\subsection{Transformation from the Jordan to the Einstein frame}\label{SM:JordanToEinstein}
We start with the Jordan frame action 
\begin{align}\label{eq:JordanFrameAction}
\mathcal{S}_{\textrm{EH+}H} &= 
\int d^4 x\sqrt{-g}\bigg[
\left(\frac{1}{2}\bar{M}_{\textrm{Pl}}^2 + \xi H^{\dag}H\right)R \nonumber \\
&+ (D_{\mu}H)^{\dag}(D^{\mu}H) -  V(H^{\dag}H)\bigg]\,,
\end{align}
that combines the Einstein-Hilbert term ($R$ is the Ricci scalar) and the Higgs sector of the SM. The Higgs potential is
$V(H^{\dag}H) = -\mu^2 H^{\dag}H + \lambda(H^{\dag}H)^2$ and $D_{\mu}H$ is the $SU(2)_L\otimes U(1)_Y$ covariant derivative acting on the Higgs doublet. For simplicity's sake, we omit the subscript indicating bare parameters and fields. The potential $V(H^{\dag}H)$ induces spontaneous symmetry  breaking, and, after expanding around the minimum of the potential, we get
\begin{align}\label{eq:SJFSSB}
\mathcal{S}_{\textrm{EH+}H} &\ni 
\int d^4 x\sqrt{-g}
\bigg\{
\frac{1}{2}\left[
\bar{M}_{\textrm{Pl}}^2 + \xi(h+v)^2
\right]R \nonumber\\
&+ \frac{1}{2}(\partial_{\mu}h)(\partial^{\mu}h) - 
\left(
\frac{1}{2}m^2h^2 + \lambda v h^3 + \frac{\lambda}{4}h^4
\right)
\bigg\}
\,,
\end{align}
where $m^2 = 2 \lambda v^2$. We only focus on the Higgs field (i.e. we work in the unitary gauge and neglect the interaction terms in the covariant derivatives). 
Next, we move to the Einstein frame.
The Lagrangian density for the Higgs field reads
 \begin{align}
\mathcal{L}_{H} = &\frac{1}{2}\left[
\frac{1}{\Omega^2} + \frac{3\bar{M}_{\textrm{Pl}}^2}{2\Omega^4}
\left(\frac{d\Omega^2}{dh}\right)^2
\right](\partial_{\mu}h)(\partial^{\mu}h) \nonumber\\
&-\frac{1}{\Omega^4}\left(
\frac{1}{2}m^2h^2 + \lambda v h^3 + \frac{\lambda}{4}h^4
\right)\,,~~~~~~~ \nonumber
 \end{align}
 with:
 \begin{equation}
     \Omega^2\equiv 1 + \frac{\xi(h+v)^2}{\bar{M}_{\textrm{Pl}}^2}\,.\label{eq:Omega}
 \end{equation}
This Lagrangian density describes a non-renormalizable scalar theory. 
This is the prize to pay, in the Einstein frame, for the non-minimal coupling of the scalar $H$ to gravity -- that is, a non-renormalizable theory -- in the Jordan frame. 
We expand the above Lagrangian up to dimension-four operators in the field $h$. We find
\begin{align}
\mathcal{L}_{H} = &\frac{1}{2}\left[
\frac{
1+\xi(1+6\xi)v^2/\bar{M}_{\textrm{Pl}}^2
}{(1+\xi v^2/\bar{M}_{\textrm{Pl}}^2)^2}
\right](\partial_{\mu}h)(\partial^{\mu}h)\nn\\
&-
\bigg[
\frac{1}{2}\frac{m^2}{(1+\xi v^2/\bar{M}_{\textrm{Pl}}^2)^2}h^2 
+ \frac{v\lambda(1-3\xi v^2\bar{M}_{\textrm{Pl}}^2)^2)}{(1+\xi v^2/\bar{M}_{\textrm{Pl}}^2)^3}h^3 \nn\\
&+ \frac{\lambda(1 - 22\xi v^2/\bar{M}_{\textrm{Pl}}^2 + 25\xi^2 v^4/\bar{M}_{\textrm{Pl}}^4)}{4(1+\xi v^2/\bar{M}_{\textrm{Pl}}^2)^4}h^4
\bigg] \nn\\
&+ \textrm{higher dim operators}\,.\label{eq:LagraXi}
\end{align} 
The field $h$ in Eq.\,(\ref{eq:LagraXi}) is not canonically  normalized. 
If we rescale
\begin{align}\label{eq:FieldRescaling}
h \to a h\,,~~~~~\textrm{with:}~~~
a\equiv \frac{1+\xi v^2/\bar{M}_{\textrm{Pl}}^2}{[1+\xi(1+6\xi)v^2/\bar{M}_{\textrm{Pl}}^2]^{1/2}}\,,
\end{align}
we find the Lagrangian density
\begin{align}
\mathcal{L}_{H} = &\frac{1}{2}(\partial_{\mu}h)(\partial^{\mu}h) - \bigg\{ 
\frac{1}{2}\frac{m^2}{[1+\xi(1+6\xi)v^2/\bar{M}_{\textrm{Pl}}^2]}h^2 \nn\\
&+ 
\frac{v\lambda(1-3\xi v^2/\bar{M}_{\textrm{Pl}}^2)}{
[1+\xi(1+6\xi)v^2/\bar{M}_{\textrm{Pl}}^2]^{3/2}
}h^3\nn \\
&+ 
\frac{\lambda(1 - 22\xi v^2/\bar{M}_{\textrm{Pl}}^2 + 25\xi^2 v^4/\bar{M}_{\textrm{Pl}}^4)}{
4[1+\xi(1+6\xi)v^2/\bar{M}_{\textrm{Pl}}^2]^{2}
}h^4\bigg\}\,,\label{eq:ModifiedHiggs}
\end{align}
with additional higher-dimensional operators that are not shown. 
In the limit $\xi\to 0$, Eq.\,(\ref{eq:ModifiedHiggs}) reproduces the Lagrangian density of the SM describing the massive Higgs excitation and its self-interactions in the broken phase of the electroweak symmetry. 
The Higgs mass is now given by 
\begin{align}
M_h^2 &\equiv \frac{m^2}{[1+\xi(1+6\xi)v^2/\bar{M}_{\textrm{Pl}}^2]} \nn\\
&= m^2\left[1 - \frac{2\xi(1+6\xi)v^2}{\bar{M}_{\textrm{Pl}}^2} + O\left(\frac{ v^4}{\bar{M}_{\textrm{Pl}}^4}\right)
\right]\,,
\end{align}
and equals the tree-level Higgs mass in the  SM (that is, $M_h^2 = m^2$) plus small gravitational corrections.
Similarly, the trilinear and qurtic Higgs couplings take the form
\begin{align}
&\textrm{trilinear:}~~\lambda v\left[
1-\frac{3}{4}\xi(5+6\xi)\frac{v^2}{\bar{M}_{\textrm{Pl}}^2}+ O\left(\frac{ v^4}{\bar{M}_{\textrm{Pl}}^4}\right)
\right]h^3\,,~~~~~~~~ \nn \\
&\textrm{quartic:}~~\frac{\lambda}{4}\left[
1-\xi(23+6\xi)\frac{v^2}{\bar{M}_{\textrm{Pl}}^2}+ O\left(\frac{ v^4}{\bar{M}_{\textrm{Pl}}^4}\right)
\right]h^4\,.
\end{align}
We now add to the Jordan frame action in  Eq.\,(\ref{eq:JordanFrameAction}) the action for a free Dirac fermion $\Psi$ with mass $M_{\Psi}$ minimally coupled to gravity and decoupled from the Higgs field $H$. 
In the Einstein frame, the key aspect is that the mass term becomes 
\begin{align}
-\frac{M_{\Psi}}{\Omega}\bar{\Psi}\Psi\,,
\end{align}
with $\Omega$ given in Eq.\,(\ref{eq:Omega}). 
If we expand $\Omega^{-1}$ and rescale the field $h$ according to Eq.\,(\ref{eq:FieldRescaling}), we find
\begin{align}
\frac{M_{\Psi}}{\Omega}\bar{\Psi}\Psi &= 
\frac{M_{\Psi}}{(1+\xi v^2/\bar{M}_{\textrm{Pl}}^2)^{1/2}}\bigg\{
1 -   \nn\\
&
\frac{\xi v h}{
\bar{M}_{\textrm{Pl}}^2
[1+\xi(1+6\xi)v^2/\bar{M}_{\textrm{Pl}}^2]^{1/2}}  \nn\\
&
- \frac{\xi(1-2\xi v^2/\bar{M}_{\textrm{Pl}}^2)}{2\bar{M}_{\textrm{Pl}}^2[1+\xi(1+6\xi)v^2/\bar{M}_{\textrm{Pl}}^2]}h^2 
+ O(h^3)
\bigg\}\bar{\Psi}\Psi\,.
\end{align}
In the Einstein frame, therefore, the $\Psi$ mass generates a tower of interaction terms that couples $\Psi$ to the Higgs field.
The leading term that survives, as expected, even if we take the limit $v \to 0$, is the effective operator
\begin{align}
\frac{\xi M_{\Psi}}{2\bar{M}_{\textrm{Pl}}^2}h^2\bar{\Psi}\Psi\,,
\end{align}
which destabilizes the Higgs mass.

\subsection{A Jordan Frame Analysis}\label{SM:JordanAnalysis}
It is possible to obtain Eq.~\eqref{eq:Higgs_mass_additive_term} working directly in the Jordan frame. For the sake of simplicity, we focus directly on terms quadratic in $h$. We further assume $g_{\mu\nu} = \eta_{\mu\nu} + \kappa \, l_{\mu\nu}$, where $\kappa \equiv 2/\bar{M}_{\textrm{Pl}}$. Since we are interested in the naturalness problem that arises in the present universe, we expand around a flat background. It is straightforward to generalized the discussion to an FRLW background, relevant for early-universe considerations. The main difference would be a shift in the pole Higgs mass due to the cosmological constant contribution, but the analysis about quantum corrections still holds.

At first order in $\kappa$, the graviton-scalar mixing induced by the non-minimal coupling is described by the term
\begin{equation}\label{eq:nonmin_vertex}
    \sqrt{-g} \, \frac{\xi}{2} R \,h^2 = \kappa \frac{\xi}{2} (l^{\mu \alpha} \partial_\mu \partial_\alpha - l \,\Box ) h^2 \, ,
\end{equation}
where $l = \eta^{\mu\nu} l_{\mu\nu}$. The overall graviton-scalar interaction can be written as
\begin{equation}
    \sqrt{-g} \mathcal{L}_H^{(1)} = \frac{\kappa}{2} l^{\mu\nu} T_{\mu\nu} \, ,
\end{equation}
where $T_{\mu\nu}$ is the effective energy-momentum tensor associated to the scalar field, i.e.
\begin{align}
    T_{\mu\nu} &= 
    -\partial_\mu h \partial_\nu h + \eta_{\mu\nu}\Big(\frac{1}{2} \partial_\alpha h \partial^\alpha h - \frac{1}{2} m^2 h^2\Big) 
    - \nonumber\\
    &\xi (\eta_{\mu\nu}\, \Box - \partial_\mu \partial_\nu) h^2 \,.
\end{align}
The corresponding amplitude is
\begin{align}
    &\frac{-i \kappa}{2} \Big[p_{1\mu} p_{2\nu} + p_{1\nu} p_{2\mu} - \eta_{\mu\nu} (p_1 \cdot p_2 - m^2) +\nonumber\\
    &2\,\xi \Big(\eta_{\mu\nu} \,q^2 - q_\mu q_\nu\Big) \Big] \,,
\end{align}
and the associated Feynman diagram is
\begin{center}

\tikzset{every picture/.style={line width=0.75pt}} 

\begin{tikzpicture}[x=0.75pt,y=0.75pt,yscale=-1,xscale=1]

\draw  [dash pattern={on 4.5pt off 4.5pt}]  (250.27,54.65) -- (250.5,148.14) ;
\draw    (250.38,98.07) .. controls (252.05,96.4) and (253.71,96.4) .. (255.38,98.07) .. controls (257.05,99.74) and (258.71,99.74) .. (260.38,98.07) .. controls (262.05,96.4) and (263.71,96.4) .. (265.38,98.07) .. controls (267.05,99.74) and (268.71,99.74) .. (270.38,98.07) .. controls (272.05,96.4) and (273.71,96.4) .. (275.38,98.07) .. controls (277.05,99.74) and (278.71,99.74) .. (280.38,98.07) .. controls (282.05,96.4) and (283.71,96.4) .. (285.38,98.07) .. controls (287.05,99.74) and (288.71,99.74) .. (290.38,98.07) .. controls (292.05,96.4) and (293.71,96.4) .. (295.38,98.07) -- (299.88,98.07) -- (299.88,98.07)(250.38,101.07) .. controls (252.05,99.4) and (253.71,99.4) .. (255.38,101.07) .. controls (257.05,102.74) and (258.71,102.74) .. (260.38,101.07) .. controls (262.05,99.4) and (263.71,99.4) .. (265.38,101.07) .. controls (267.05,102.74) and (268.71,102.74) .. (270.38,101.07) .. controls (272.05,99.4) and (273.71,99.4) .. (275.38,101.07) .. controls (277.05,102.74) and (278.71,102.74) .. (280.38,101.07) .. controls (282.05,99.4) and (283.71,99.4) .. (285.38,101.07) .. controls (287.05,102.74) and (288.71,102.74) .. (290.38,101.07) .. controls (292.05,99.4) and (293.71,99.4) .. (295.38,101.07) -- (299.88,101.07) -- (299.88,101.07) ;
\draw  [fill={rgb, 255:red, 0; green, 0; blue, 0 }  ,fill opacity=1 ] (250.3,63.64) -- (251.8,70.06) -- (248.8,70.06) -- cycle ;
\draw  [fill={rgb, 255:red, 0; green, 0; blue, 0 }  ,fill opacity=1 ] (248.56,99.57) .. controls (248.56,98.57) and (249.38,97.75) .. (250.38,97.75) .. controls (251.39,97.75) and (252.2,98.57) .. (252.2,99.57) .. controls (252.2,100.58) and (251.39,101.39) .. (250.38,101.39) .. controls (249.38,101.39) and (248.56,100.58) .. (248.56,99.57) -- cycle ;
\draw  [fill={rgb, 255:red, 0; green, 0; blue, 0 }  ,fill opacity=1 ] (250.3,123.64) -- (251.8,130.06) -- (248.8,130.06) -- cycle ;

\draw (229.99,43.05) node [anchor=north west][inner sep=0.75pt]    {$p_{2}$};
\draw (229.99,133.58) node [anchor=north west][inner sep=0.75pt]    {$p_{1}$};
\draw (303.33,94.59) node [anchor=north west][inner sep=0.75pt]    {$q$};
\draw (254.47,83.27) node [anchor=north west][inner sep=0.75pt]  [font=\footnotesize,rotate=-0.57]  {$\mu \nu $};

\end{tikzpicture}

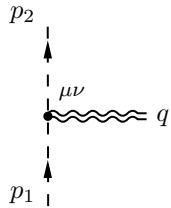
\captionof{figure}{\em Feynman diagram for the graviton-scalar vertex. }
\end{center}

The fact that the non-minimal interaction sources a contribution that involves only $q_\mu$, the graviton momentum, is understood by noticing that the derivatives in Eq.~\eqref{eq:nonmin_vertex} can be conveniently carried on $l_{\mu\nu}$, up to irrelevant boundary terms.

The only diagrams that gives a contribution $\sim M_\Psi^4 / \bar{M}_{\textrm{Pl}}^2$ is the following tadpole

\begin{center}

\tikzset{every picture/.style={line width=0.75pt}} 

\begin{tikzpicture}[x=0.75pt,y=0.75pt,yscale=-1,xscale=1]

\draw  [dash pattern={on 4.5pt off 4.5pt}]  (261.97,97.39) -- (151.36,97.15) ;
\draw    (207.75,98.21) .. controls (206.11,96.52) and (206.14,94.85) .. (207.83,93.21) .. controls (209.52,91.57) and (209.55,89.9) .. (207.92,88.21) .. controls (206.29,86.52) and (206.32,84.85) .. (208.01,83.21) .. controls (209.7,81.57) and (209.73,79.9) .. (208.1,78.21) .. controls (206.46,76.52) and (206.49,74.85) .. (208.18,73.21) .. controls (209.87,71.57) and (209.9,69.9) .. (208.27,68.21) .. controls (206.64,66.52) and (206.67,64.85) .. (208.36,63.21) .. controls (210.05,61.58) and (210.08,59.91) .. (208.44,58.22) -- (208.47,56.53) -- (208.47,56.53)(210.75,98.26) .. controls (209.11,96.57) and (209.14,94.9) .. (210.83,93.26) .. controls (212.52,91.62) and (212.55,89.95) .. (210.92,88.26) .. controls (209.29,86.57) and (209.32,84.9) .. (211.01,83.26) .. controls (212.7,81.62) and (212.73,79.95) .. (211.09,78.26) .. controls (209.46,76.57) and (209.49,74.9) .. (211.18,73.27) .. controls (212.87,71.63) and (212.9,69.96) .. (211.27,68.27) .. controls (209.64,66.58) and (209.67,64.91) .. (211.36,63.27) .. controls (213.05,61.63) and (213.08,59.96) .. (211.44,58.27) -- (211.47,56.58) -- (211.47,56.58) ;
\draw  [fill={rgb, 255:red, 0; green, 0; blue, 0 }  ,fill opacity=1 ] (244.23,97.38) -- (236.63,99.08) -- (236.65,95.62) -- cycle ;
\draw  [fill={rgb, 255:red, 0; green, 0; blue, 0 }  ,fill opacity=1 ] (209.24,99.18) .. controls (208.06,99.18) and (207.09,98.24) .. (207.09,97.08) .. controls (207.1,95.92) and (208.06,94.98) .. (209.25,94.98) .. controls (210.44,94.98) and (211.4,95.92) .. (211.4,97.08) .. controls (211.4,98.24) and (210.43,99.18) .. (209.24,99.18) -- cycle ;
\draw  [fill={rgb, 255:red, 0; green, 0; blue, 0 }  ,fill opacity=1 ] (181.28,97.36) -- (173.68,99.06) -- (173.7,95.6) -- cycle ;
\draw   (193.04,40.04) .. controls (193.04,30.91) and (200.62,23.52) .. (209.97,23.52) .. controls (219.32,23.52) and (226.9,30.91) .. (226.9,40.04) .. controls (226.9,49.16) and (219.32,56.56) .. (209.97,56.56) .. controls (200.62,56.56) and (193.04,49.16) .. (193.04,40.04) -- cycle ;
\draw  [fill={rgb, 255:red, 0; green, 0; blue, 0 }  ,fill opacity=1 ] (209.97,58.66) .. controls (208.78,58.65) and (207.82,57.71) .. (207.82,56.55) .. controls (207.82,55.39) and (208.79,54.45) .. (209.97,54.46) .. controls (211.16,54.46) and (212.13,55.4) .. (212.12,56.56) .. controls (212.12,57.72) and (211.16,58.66) .. (209.97,58.66) -- cycle ;

\draw (215.83,68.73) node [anchor=north west][inner sep=0.75pt]    {$q$};
\draw (168.45,102.77) node [anchor=north west][inner sep=0.75pt]    {$p$};
\draw (240.45,102.77) node [anchor=north west][inner sep=0.75pt]    {$p$};

\end{tikzpicture}

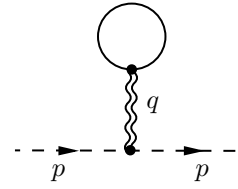
\captionof{figure}{\em Tadpole contribution to the Higgs propagator.}
\label{fig:tadpole}
\end{center}
Due to momentum conservation only the zero graviton modes mediate this tadpole, leading to an infrared divergence. We regularize this object by assuming $q^2 \neq 0$ and subsequently taking the limit $q^2 \to 0$.

The contribution to the Higgs-self energy is then
\begin{align}\label{eq:self_energy_Higgs}
    i \Sigma_{grav} = \frac{i\kappa^2}{2 (4\pi)^2} \frac{1}{q^2} M_\Psi^2 A_0(M_\Psi^2) \Big( 3 \xi q^2 + m_H^2 \Big) \,.
\end{align}

As expected, the contribution coming from the standard vertex, obtained in the minimally coupled case, is proportional to $m_H^2$ and thus vanishes if we send $m_H^2 \to 0$. On the contrary the contribution from the non-minimal coupling cancels the soft graviton momentum at the denominator, leaving a finite contribution for $q^2 \to 0$, and does not vanish in the $m_H^2 \to 0$ limit. Thus, the additive contribution to the running Higgs mass is again
\begin{equation}\label{eq:Higgs_additive_correction}
    \delta m_H^2 \sim \xi \frac{1}{(4\pi)^2} \frac{M_\Psi^4}{\bar{M}_{\textrm{Pl}}^2} \,
\end{equation}
in agreement with Eq.~\eqref{eq:Higgs_mass_additive_term}.


Finally, we point out that an alternative way to cure the infrared divergence in Fig.~\ref{fig:tadpole} is to use a cosmological constant. 
Indeed, an additional term $\sqrt{-g} \, \Lambda$ gives at fist order the following vertex proportional to $\eta_{\mu\nu}$
\begin{center}
\tikzset{every picture/.style={line width=0.75pt}} 

\begin{tikzpicture}[x=0.75pt,y=0.75pt,yscale=-1,xscale=1]

\draw    (150.1,48.67) .. controls (151.76,46.99) and (153.43,46.98) .. (155.1,48.64) .. controls (156.77,50.3) and (158.44,50.29) .. (160.1,48.62) .. controls (161.76,46.95) and (163.43,46.94) .. (165.1,48.59) .. controls (166.77,50.25) and (168.44,50.24) .. (170.1,48.57) .. controls (171.76,46.9) and (173.43,46.89) .. (175.1,48.54) .. controls (176.77,50.2) and (178.44,50.19) .. (180.1,48.52) .. controls (181.76,46.85) and (183.43,46.84) .. (185.1,48.49) .. controls (186.77,50.15) and (188.44,50.14) .. (190.1,48.47) .. controls (191.76,46.8) and (193.43,46.79) .. (195.1,48.44) .. controls (196.77,50.1) and (198.44,50.09) .. (200.1,48.42) -- (200.6,48.42) -- (200.6,48.42)(150.12,51.67) .. controls (151.78,49.99) and (153.45,49.98) .. (155.12,51.64) .. controls (156.79,53.3) and (158.46,53.29) .. (160.12,51.62) .. controls (161.78,49.95) and (163.45,49.94) .. (165.12,51.59) .. controls (166.79,53.25) and (168.46,53.24) .. (170.12,51.57) .. controls (171.78,49.9) and (173.45,49.89) .. (175.12,51.54) .. controls (176.79,53.2) and (178.46,53.19) .. (180.12,51.52) .. controls (181.78,49.85) and (183.45,49.84) .. (185.12,51.49) .. controls (186.79,53.15) and (188.46,53.14) .. (190.12,51.47) .. controls (191.78,49.8) and (193.45,49.79) .. (195.12,51.44) .. controls (196.79,53.1) and (198.46,53.09) .. (200.12,51.42) -- (200.62,51.42) -- (200.62,51.42) ;
\draw   (185.07,49.92) .. controls (185.07,41.6) and (192.03,34.86) .. (200.61,34.86) .. controls (209.2,34.86) and (216.16,41.6) .. (216.16,49.92) .. controls (216.16,58.23) and (209.2,64.97) .. (200.61,64.97) .. controls (192.03,64.97) and (185.07,58.23) .. (185.07,49.92) -- cycle ; \draw   (189.62,39.27) -- (211.6,60.56) ; \draw   (211.6,39.27) -- (189.62,60.56) ;
\draw  [fill={rgb, 255:red, 0; green, 0; blue, 0 }  ,fill opacity=1 ] (200.61,52.02) .. controls (199.42,52.02) and (198.46,51.07) .. (198.46,49.91) .. controls (198.46,48.75) and (199.42,47.81) .. (200.61,47.82) .. controls (201.8,47.82) and (202.77,48.76) .. (202.76,49.92) .. controls (202.76,51.08) and (201.8,52.02) .. (200.61,52.02) -- cycle ;

\end{tikzpicture}

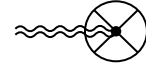
\captionof{figure}{\em Counterterm generated by the addition of a cosmological constant.}
\end{center}
Thus, the constant $\Lambda$ can be fine-tuned in such a way that the tadpole (without including the graviton propagator) vanishes. However, this works only in the minimal case, since $\Lambda$ can cancel only a constant quantity and not a momentum-dependent one. Therefore, the contribution proportional to $\xi$ in Eq.~\eqref{eq:self_energy_Higgs} will unavoidably survive.

\subsection{Inflaton dynamics and relation to cosmological observables}
For the sake of clarity, let us take advantage of this point to define our notation. 
The Hubble rate is $H\equiv \dot{a}/a$, 
with $\dot{} \equiv d/dt$ where $t$ is the cosmic time  and $a$ the scale factor of the flat Friedmann-Lema\^{\i}tre-Robertson-Walker (FLRW) metric $ds^2 = dt^2 - a^2(t)d\vec{x}^2$, with $\vec{x}$ comoving coordinates. 
The $e$-fold time $N$ is defined by $dN=Hdt$.  
The Hubble-flow parameters $\epsilon_{i}$ (for $i\geqslant 1$) are defined by the recursive relation
\begin{align}
	\epsilon_{i} \equiv \frac{\dot{\epsilon}_{i-1}}{H\epsilon_{i-1}}\,,~~~~~\textrm{with:}~~~
	\epsilon_0 \equiv \frac{1}{H}\,.\label{eq:HubblePar1}
\end{align}
As customary, we simply indicate as $\epsilon$ the first Hubble parameter, $\epsilon \equiv \epsilon_1 = -\dot{H}/H^2$. 
Instead of the second Hubble parameter $\epsilon_2$, sometimes it is useful to introduce the Hubble parameter $\eta$ 
defined by
\begin{align}
	\eta \equiv - \frac{\ddot{H}}{2H\dot{H}} 
	= \epsilon - 
	\frac{1}{2}\frac{d\log\epsilon}{dN}\,,~~~~
	\textrm{with:}~~~\epsilon_2 = 2\epsilon - 2\eta\,.\label{eq:HubblePar2}
\end{align}
Using the number of $e$-folds as time variable,  the inflaton equation of motion reads
\begin{align}
	\frac{d^2\phi}{dN^2} + \left[3 - \frac{1}{2\MPl^2}\left(\frac{d\phi}{dN}\right)^2\right]
	\left[\frac{d\phi}{dN} + \MPl^2\frac{d\log V(\phi)}{d\phi}
	\right] = 0\,,\label{eq:EoM}
\end{align}
and, in turn, the Hubble parameters take the form
\begin{align}
	\epsilon  = \frac{1}{2\MPl^2}\left(\frac{d\phi}{dN}\right)^2\,,~~~\eta = 
	3 + \frac{\MPl^2(3 - \epsilon)V^{\prime}(\phi)}{V(\phi)(d\phi/dN)}
	\,,
\end{align}
while the Hubble rate is related to the inflaton potential by means of the Friedmann equation
\begin{align}
	(3-\epsilon)H^2 = V(\phi)\,.
\end{align}

We indicate the slow-roll approximation of $\epsilon$ and $\eta+\epsilon$ as $\epsilon_V$ and $\eta_V$. We have 
\begin{align}\label{eq:slow_roll_parameters}
	\epsilon_V(\phi) = 
	\frac{\MPl^2}{2}\left(\frac{V^{\prime}(\phi)}{V(\phi)}\right)^2\,,~~~
	\eta_V(\phi) = \MPl^2\frac{V^{\prime\prime}(\phi)}{V(\phi)}\,.
\end{align}
As is well-known, the comparison of the theory's predictions with cosmological observables in the slow-roll limit consists of three steps.
\begin{itemize}
	\item[$\ast$] By imposing $\epsilon = 1$, we find the field value at the end of the accelerated phase of inflationary expansion, $\phi_{\textrm{end}}$. 
	\item[$\ast$] Integrating $dN = Hdt$ in field space, we find the duration of inflation in the interval $(\phi_{\textrm{end}},\phi_{*})$, where we indicate with $\phi_{*}$ the field value at which the long-scale CMB modes exited the Hubble horizon during inflation. In the slow-roll approximation, we have 
	\begin{center}
	   \begin{equation}
		N = -\frac{1}{\MPl^2}\int_{\phi_{*}}^{\phi_{\textrm{end}}}
		\frac{V(\phi)}{V^{\prime}(\phi)}d\phi\,.\label{eq:EfoldTime}
	\end{equation}\\[0.3ex] 
	\end{center}
    By requiring $N \in (50,60)$ $e$-folds of inflation, we extract $\phi_{*}$ using Eq.\,(\ref{eq:EfoldTime}).
	\item[$\ast$] Finally, we compute 
	\begin{align}
		n_s = 1 + 2\eta_V(\phi_*) - 6\epsilon_V(\phi_*)\,,~~~~
		r = 16\epsilon_V(\phi_*)\,,
	\end{align}
	and compare with cosmological data.
\end{itemize}

\end{document}